\documentclass{ws-rv975x65}  

\begin{document}


\centerline{ANOMALIES TO ALL ORDERS}
\bigskip

\markboth{S. Adler}{Anomalies to All Orders}

\author{Stephen L. Adler}

\address{Institute for Advanced Study\\
Einstein Drive, Princeton, NJ 08540 USA\\
E-mail: adler@ias.edu}

\begin{abstract} 
I give an account of my involvement with the chiral anomaly, and  
with the nonrenormalization 
theorem for the chiral anomaly and the all orders calculation of  
the trace anomaly, as well as related 
work by others.   I then briefly discuss implications of 
these results for more recent developments in anomalies in supersymmetric 
theories.  
\end{abstract}

\section{Introduction}
A distinguishing feature of the chiral and trace anomalies in gauge theory 
is that their coefficients can be determined to all orders of perturbation 
theory.  My aims in this article are to survey the historical 
route leading to the discovery of this property of anomalies, and to review 
its current status.  In the first part I discuss the nonrenormalization of 
the chiral anomaly, in its historical context and then from a modern 
perspective.  In the second part I discuss the closely analogous results 
linking the coefficient of the trace anomaly to the renormalization group 
$\beta$ function.  In the third part, I conclude by 
discussing the interplay of these 
two results in the context of supersymmetric theories. 

\section{Nonrenormalization of the Chiral Anomaly \label{NCA}}

\subsection{Setting the Stage \label{stage}}

My 1969 paper on anomalies [\refcite{adler69a}] consisted of two parts.  
The first was a 
discussion of the axial anomaly in spinor electrodynamics, and represented  
work done during the spring and summer of 1968 and written up in longhand 
form (awaiting typing on my return to Princeton) while I was at the Aspen 
Center of Physics.  The second part consisted of an Appendix  and additional 
footnotes,   
written after Sidney Coleman arrived in Aspen towards the end of my stay 
there, and told me about the independent work done by Bell and Jackiw [\refcite{belljack}]  
on the anomaly in the context of the partially conserved axial current (PCAC) 
calculation of $\pi^0 \to 2 \gamma$ decay in the sigma model.  
Both parts are relevant to the 
story of the anomaly nonrenormalization theorem, and so I shall begin by 
discussing them in some detail.  

I got into the subject of anomalies in an indirect way, through exploration 
during 1967--1968 of the speculative idea that the muon--electron mass 
difference could be accounted for by giving the muon an additional magnetic 
monopole electromagnetic coupling through an axial-vector current, which 
somehow was nonperturbatively renormalized to zero.  After much fruitless 
study of the integral equations for the axial-vector vertex part, I decided 
in the spring of 1968 to first try to answer a well-defined question, which 
was whether the axial-vector vertex in QED was renormalized by multiplication 
by $Z_2$, as I had been implicitly assuming.  At the time when I turned 
to this question, I had just started a 6-week visit to 
the Cavendish Laboratory in 
Cambridge, England after flying to London with my family 
on April 21, 1968.\footnote{These details of dates were 
recorded by my ex-wife Judith in my oldest daughter 
Jessica's ``baby book''.}  In the Cavendish I   
shared an office 
with my former adviser Sam Treiman, and was 
enjoying the opportunity to try a new project not requiring extensive 
computer analysis; I had only a month before finished my Annals of 
Physics paper [\refcite{adler68}] on weak pion production, which had required  
extensive computation, not easy to do in those days when one had to wait 
hours or even a day for the results of a computer run.  

My interest in the multiplicative  
renormalization question had been piqued by work of 
van Nieuwenhuizen, in which he had attempted to demonstrate the 
finiteness to all orders of radiative corrections to $\mu$-decay, using 
an argument based on subtraction of renormalization constants that I knew  
to be incorrect beyond leading order.  I had learned about this work during  
the previous summer, when I was a lecturer at the Varenna summer school held 
by Lake Como from July 17-29, 1967, at which van Nieuwenhuizen had given a  
seminar on this topic that was critiqued 
by Bjorken, another lecturer.\footnote{I wish to thank Peter van Nieuwenhuizen for a phone conversation clarifying this part of the history.  In my 1998 Dirac lecture [\refcite{adler98}], 
and several archival historical accounts, I had written ``My interest in the 
multiplicative renormalization question had been piqued by a preprint I 
received from Peter van Nieuwenhuizen, in which he attempted to show that 
the axial vector vertex is made finite by the usual renormalizations, using 
an argument based on subtractive renormalization that I saw was incorrect.''
On reexamining my files in the course of preparing this article, 
I found that my copy of van Nieuwenhuizen's 
undated preprint entitled ``Finiteness of radiative corrections in all 
orders to $\mu$-decay'' had been sent, in response to a letter I wrote 
to Veltman, by Veltman's secretary Ms. Rietveld, who wrote a 
cover letter  dated May 28, 1968 expressing the hope that the preprint would 
reach me while I was still in England. Thus I could not have  
seen the preprint until the end of my stay in Cambridge. I do not have   
a copy of my letter to Veltman inquiring about a preprint version 
of van Nieuwenhuizen's talk, to which Ms. Rietveld's letter was a response.}
Working in the old Cavendish, 
I rather rapidly found an inductive multiplicative renormalizability proof, 
paralleling the one in Bjorken and Drell  [\refcite{bjor-drell}] for finiteness of $Z_2$ 
times the vector vertex.  I prepared a detailed outline for a paper 
describing the proof, but before writing things up, I decided as a check 
to test whether the formal argument for the closed loop part of the Ward 
identity  worked in the case of the smallest loop diagram.  
This is a triangle diagram with one axial and two vector vertices 
\big(the $AVV$ triangle; see Fig. 1(a)\big),   
which because of Furry's theorem ($C$ invariance) has no analog in the 
vector vertex case.  I knew from a student seminar that I had 
attended during my graduate study at Princeton that this diagram had been 
explicitly calculated using a gauge-invariant regularization 
by Rosenberg  [\refcite{rose}], who was interested in the 
astrophysical process $\gamma_V+ \nu \to \gamma + \nu$, with $\gamma_V$ a 
virtual photon emitted by a nucleus.  I got Rosenberg's paper, tested 
the Ward identity, and to my astonishment (and Treiman's when I told him 
the result) found that it failed!  I soon found that the problem was that 
my formal proof used a shift of integration variables inside a linearly 
divergent integral, which (as I again recalled from student reading) had 
been analyzed in an Appendix to the classic text of Jauch and Rohrlich 
 [\refcite{jauch55}], with a calculable constant remainder.  For all closed 
loop contributions to the axial vertex 
in Abelian electrodynamics with larger numbers of vector vertices 
\big(the $AVVVV$, $AVVVVVV$,... loops; see Fig. 1(b)\big),  
the fermion loop integrals for fixed photon momenta are highly 
convergent and the shift of integration variables needed in the Ward 
identity {\it is} valid, so proceeding in 
this fermion loop-wise fashion there were  
apparently no further additional or ``anomalous'' contributions to the   
axial-vector Ward identity.  With this fact in the back of my mind 
I was convinced from the outset that the anomalous contribution to the 
axial Ward identity would come just from the triangle diagram, with no 
renormalizations of the anomaly coefficient arising 
from higher order $AVV$ diagrams with virtual photon insertions.  
\begin{figure}[th]		
\centerline{\psfig{file=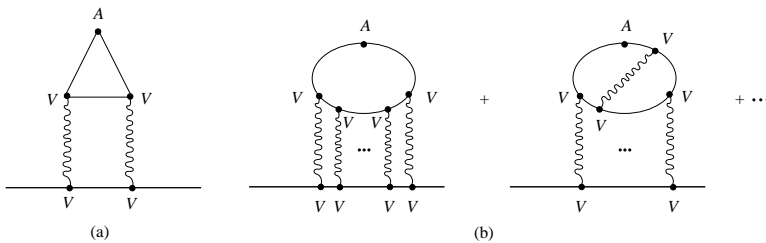,width=4in}}
\vspace*{8pt}
\caption{Fermion loop diagram contributions to the axial-vector vertex  
part.  Solid lines are fermions, and dashed lines are photons.  (a) The 
smallest loop, the $AVV$ triangle diagram.  (b)  Larger loops with four 
or more vector vertices, which (when summed over vertex orderings) 
obey normal Ward identities. \label{fig1a-b}}
\end{figure}


In early June, at the end of my 6 weeks in Cambridge,  
I returned to the US and then went to Aspen, where 
I spent the summer working out a manuscript on the properties of the 
axial anomaly, which became the body (pages 2426-2434) of the final 
published version [\refcite{adler69a}].  Several of the things done there 
will figure  
in the further discussion of anomaly nonrenormalization.  The 
first was a calculation of the field theoretic form of the anomaly,
giving the now well-known result 
\begin{equation}
\partial^{\mu} j_{\mu}^5(x) = 2im_0j^5(x) 
+ {\alpha_0 \over 4 \pi} 
F^{\xi \sigma}(x) F^{\tau \rho}(x) \epsilon_{\xi\sigma \tau \rho} ~~~,
\label{aid}
\end{equation}
with $j_{\mu}^5=\overline \psi \gamma_{\mu}\gamma_5 \psi$ 
the axial-vector current 
(referred to above as $A$), 
$j^5=\overline \psi \gamma_5 \psi$ the pseudoscalar 
current, and with $m_0$ and $\alpha_0$ the 
(unrenormalized) fermion mass and coupling constant.  The second was  
a demonstration that because of the anomaly, $Z_2$ is no longer the 
multiplicative renormalization constant for the axial-vector vertex, as a   
result of the diagram drawn in Fig.~1(a) in which the $AVV$ triangle 
is joined  to an electron line with two virtual photons.   
Instead, the axial-vector vertex is made finite by multiplication by 
the renormalization constant 
\begin{equation}
Z_A=Z_2[1+{3\over 4}(\alpha_0/\pi)^2 \log(\Lambda^2/m^2)+...] ~~~,
\label{bit}
\end{equation}
thus giving an answer to the question with which I started my investigation.
Thirdly, as an application of this result, I showed that the anomaly leads,  
in fourth order of perturbation theory, to 
infinite radiative corrections to the current-current theory of 
$\nu_{\mu} \mu$ and $\nu_{e}e$ scattering, but that this 
infinity can be cancelled between   
different fermion species by adding appropriate $\nu_{\mu} e$ and 
$\nu_e \mu$ scattering terms to the Lagrangian.  This result is related  
to the fact, also discussed in my paper, that the asymptotic behavior of 
the $AVV$ triangle diagram saturates the bound given by the Weinberg power 
counting rules, rather than being one power better as is the case for  
the $AVVVV$ and higher loop diagrams.  Finally, I also showed that a gauge 
invariant chiral generator still exists in the presence of the anomaly.   
Although not figuring in our subsequent discussion here, in its non-Abelian 
generalization this was relevant (as reviewed in Coleman [\refcite{cole}]) to later 
discussions of the $U(1)$ problem in quantum chromodynamics (QCD), leading     
up to the solution given by 't Hooft [\refcite{thooft}].

No sooner was this part of my paper completed than  
Sidney Coleman arrived in Aspen from Europe, and told me that Bell and 
Jackiw (published as Bell and Jackiw, [\refcite{belljack}]) 
had independently discovered the anomalous behavior of the $AVV$ 
triangle graph, in the context of a sigma model investigation of the
Veltman [\refcite{velt67}]--Sutherland [\refcite{dgs67}] theorem stating that 
$\pi^0 \to \gamma \gamma$ decay is forbidden in a PCAC calculation.  
The Sutherland--Veltman theorem is a kinematic statement 
about the $AVV$ three-point function, which asserts 
that if the momenta associated 
with the currents $A,V,V$ are respectively $q,k_1,k_2$, then the requirement 
of gauge invariance on the vector currents forces the $AVV$ vertex to be 
of order $q k_1 k_2$ in the external momenta.  Hence when one applies a 
divergence to the axial-vector vertex and uses the standard PCAC relation 
(with the quark current ${\cal F}_{3\mu}^5$ the 
analog of ${1\over 2}j_{\mu}^5$ )
\begin{equation}
\partial^{\mu} {\cal F}_{3\mu}^5(x)= 
(f_{\pi}/\sqrt{2}) \phi_{\pi}(x)~~~,
\label{cab}
\end{equation}
with $\phi_{\pi}$ the pion field and $f_{\pi}$ the charged pion decay 
constant, one finds that the $\pi^0 \to \gamma \gamma$ 
matrix element is of order $q^2 k_1 k_2$, and hence vanishes in the 
soft pion limit $q^2 \to 0$. Bell and Jackiw analyzed this result by  
a perturbative calculation in the $\sigma$-model, in 
which PCAC is formally built 
in from the outset, and found a non-vanishing result for the 
$\pi^0 \to \gamma \gamma$ amplitude, which they traced back to the fact 
that the regularized $AVV$ triangle diagram cannot be defined to satisfy 
the requirements of both PCAC and gauge invariance. This constituted the  
``PCAC Puzzle'' referred to in the title of their paper.  They then proposed 
to modify the original $\sigma$-model by adding further regulator fields 
with mass-dependent coupling constants 
in such a manner as to simultaneously enforce gauge invariance and PCAC,  
thus enforcing the Sutherland-Veltman prediction of a vanishing 
$\pi^0 \to \gamma \gamma$ decay amplitude.  In the words of Bell and 
Jackiw in their paper,  ``It has to be insisted that the introduction of 
this mass dependence of coupling constants is not an arbitrary step in the 
PCAC context.  If a regularization is introduced to define the theory, it 
must respect any formal properties which are to be appealed to.''  And 
again in concluding their paper, they stated ``To the complaint that we have 
changed the theory, we answer that only the revised version embodies 
simultaneously the ideas of PCAC and gauge invariance.''

It was immediately clear to me, in the course of the conversation with 
Sidney Coleman, that introducing additional regulators to eliminate 
the anomaly would entail renormalizability problems in $\sigma$ meson 
scattering, and was not the correct way to proceed.  However, it was also 
clear that Bell and Jackiw had made an important observation in tying the 
anomaly to the Sutherland--Veltman theorem for $\pi^0 \to \gamma \gamma$ 
decay, and that I could use the sigma-model version of Eq.~(\ref{aid}) to get 
a nonzero prediction for the $\pi^0 \to \gamma \gamma$ amplitude, with 
the whole decay amplitude 
arising from the anomaly term.  I then wrote an Appendix 
to my paper (pages 2434-2438), clearly delineated from the manuscript that 
I had finished before Sidney's arrival, in which I gave a detailed rebuttal 
of the regulator construction, by showing that the anomaly could not be 
eliminated without spoiling either gauge-invariance or 
renormalizability. \big(In later discussions I added unitarity to this 
list, to exclude the possibility of canceling the anomaly by adding 
a term to the axial current 
with a $\partial_{\mu}/(\partial_{\lambda})^2 $ singularity.\big)  
In this Appendix I also used an anomaly 
modified PCAC equation   
\begin{equation}
\partial^{\mu} {\cal F}_{3\mu}^5(x)= (f_{\pi}/\sqrt{2}) \phi_{\pi}(x)
+S {\alpha_0 \over 4 \pi} 
F^{\xi \sigma}(x) F^{\tau \rho}(x) \epsilon_{\xi\sigma \tau \rho}~~~,
\label{dip}
\end{equation}   
with $S$ a constant determined by the constituent fermion charges and 
axial-vector couplings, to obtain a PCAC formula for the 
$\pi^0 \to \gamma \gamma$ amplitude $F$ (with $\mu_{\pi}$ the pion mass)
\begin{equation}
F=-(\alpha/\pi) 2S \sqrt{2} \mu_{\pi}^2/f_{\pi}~~~.
\label{edt}
\end{equation}
Although the axial anomaly, in the context of breakdown 
of the ``pseudoscalar-pseudovector equivalence theorem'', had in fact been 
observed much earlier, starting with Fukuda and Miyamoto [\refcite{fukuda49}] and 
Steinberger [\refcite{stein49}] and continuing to Schwinger [\refcite{sch51}], my paper broke new   
ground by treating 
the anomaly neither as a baffling calculational result, nor as  
a field theoretic artifact to be 
eliminated by a suitable regularization scheme, but instead as a real 
physical effect (breaking of classical symmetries by the quantization 
process)  with observable physical consequences.  

This point of view was not immediately embraced by everyone else.    
After completing my Appendix I sent Bell and Jackiw copies of my longhand 
manuscript, and an interesting correspondence ensued.  In a letter dated 
August 25, 1968, Jackiw was skeptical whether one could extract concrete 
physical predictions from the anomaly, and whether one could 
augment the divergence of the axial-vector current by a definite 
extra electromagnetic contribution, as in the modified PCAC equation of 
Eq.~(\ref{dip}).   
Bell, who was traveling, wrote me on Sept. 2, 
1968, and was more appreciative of the possibility of using a modified 
PCAC to get a formula for the neutral pion decay amplitude, writing  
``The general idea of adding some quadratic electromagnetic 
terms to PCAC has been in our
minds since Sutherland's $\eta$ problem.  We did not see what to do with 
it.''  
He also defended the approach he and Jackiw had taken, writing 
``The reader may be left with the impression that your development is 
contradictory to ours, rather than complementary.  Our first observation 
is that the $\sigma$ model interpreted in a conventional way just does not
have PCAC.  This is already a resolution of the puzzle, and the one which 
you develop in a very nice way.  We, interested in the $\sigma$-model only 
as exemplifying PCAC, choose to modify the conventional procedures, in order 
to exhibit a model in which general PCAC reasoning could be illustrated 
in explicit calculation.''
In recognition of this letter from John Bell, whom I revered, I added a 
footnote 15 to my manuscript saying ``Our results do not 
contradict those of Bell and Jackiw, but rather complement them.  The main 
point of Bell and Jackiw is that the $\sigma$ model interpreted in the 
conventional way, does not satisfy the requirements of PCAC.  Bell and 
Jackiw modify the $\sigma$ model in such a way as to restore PCAC.  We, on 
the other hand, stay within the conventional $\sigma$ model, and try to 
systematize and exploit the PCAC breakdown.''  This footnote, which 
contradicts statements made in the text of my paper, has puzzled 
a number of people; in retrospect, rather than writing it as a paraphrase 
of Bell's words, I should have quoted directly from Bell's letter.  

Following this correspondence, my paper was typed on my return to Princeton 
and was received by Physical Review on Sept. 24, 1968.  (Bell and Jackiw's 
paper [\refcite{belljack}], a CERN preprint dated July 16, 1968,  
was submitted to Il Nuovo Cimento, and received by that journal 
on Sept. 11, 1968).  My paper was accepted along with a signed referee's 
report from Bjorken, stating 
``This paper opens a topic similar to the old controversies on photon mass 
and nature of vacuum polarization.  The lesson there, as I (no doubt 
foolishly) predict will happen 
here, is that infinities in diagrams are really 
troublesome, and that if the cutoff which is used violates a cherished 
symmetry of the theory, the results do not respect the symmetry.  I will 
also predict a long chain of papers devoted to the question the author has 
raised, culminating in a clever renormalizable cutoff which respects chiral 
symmetry and which, therefore, removes Adler's extra term.''
Thus,   acceptance of the point of view that I had advocated was 
not immediate, but only followed over time.  In 1999, 
Bjorken was a speaker at my 60th birthday conference at the 
Institute for Advanced 
Study, and amused the audience by reading from his report, and then very 
graciously gave me his file copy, with an appreciative inscription, as a 
souvenir.  

The viewpoint that the anomaly determined the $\pi^0 \to \gamma \gamma$ 
decay amplitude had significant physical consequences.  In the Appendix 
to my paper, I showed that the value of the parameter $S$ implied by the 
fractionally charged quark model gave a decay amplitude that was roughly 
a factor of 3 too small, whereas assuming integrally charged quarks gave 
an amplitude that agreed, within the expected accuracy of PCAC, with 
experiment.  In a conference talk a year later, in September 1969 
[\refcite{adler70a}] I reviewed the situation, and noted that the integrally 
charged triplet model of Han and Nambu [\refcite{han-nam}] \big(see also  
Tavkhelidze [\refcite{tavk65}] \big) also agreed with 
the experimental neutral pion decay amplitude.  These were the first 
indications that neutral pion decay provides empirical evidence for what 
we now call the ``color'' degree of freedom of the strong interactions.
For recent Archive reprintings of some of the seminal papers on color, see 
Bardeen, Fritzsch, and Gell-Mann (hep-ph/0211388) and Fritzsch and Gell-Mann 
(hep-ph/0301127).  Standard references for anomaly physics as of 1970 are 
my Brandeis lectures [\refcite{adler70b}]  and Jackiw's Brookhaven lectures [\refcite{jack70}].  

Before leaving the subject of the early history of the anomaly and its 
antecedents, perhaps this is the appropriate place to mention the paper 
of Johnson and Low [\refcite{jklow}], which showed that the 
Bjorken [\refcite{bjor66}] --Johnson--Low [\refcite{jklow}] (BJL) method of 
identifying formal commutators with an 
infinite energy limit of Feynman diagrams gives, in significant cases, 
results that differ from the naive field-theoretic evaluation of these 
commutators.  This method was later used by Jackiw and Johnson [\refcite{jackjohn}] and by 
Boulware and myself \big(Adler and Boulware [\refcite{adler-boulware}]\big) 
to show that the $AVV$ axial anomaly can be reinterpreted 
in terms of anomalous commutators.  This line of investigation, however, 
did not readily lend itself to a determination of anomaly effects beyond 
leading order.  For example, I still have in my files an unpublished 
manuscript (circa 1966) attempting to use the  BJL method to tackle 
a simpler problem, that of proving that the Schwinger 
term in quantum electrodynamics (QED) 
is a $c$-number to all orders of perturbation 
theory. I believe that this result is true (and it may well have been  
proved by now using operator product expansion methods), 
but I was not able at that time to achieve 
sufficient control of the BJL limits of high order diagrams with general  
external legs to give a proof.

\subsection{Anomaly Nonrenormalization \label{anom-non}}    

We are now ready to turn to our main story, the determination   
of anomalies beyond leading order in perturbation theory. 
Before the neutral pion low energy theorem could be used as evidence for 
the charge structure of quarks, one needed to be sure that there were no 
perturbative corrections to the anomaly and the low energy theorem 
following from it.  As I noted above, the 
fermion loop-wise argument that I used 
in my  original treatment left me convinced that only the lowest order 
$AVV$ diagram would contribute to the anomaly, but this was not a proof. 
This point of view was challenged in the article by Jackiw and Johnson 
[\refcite{jackjohn}], received by the Physical Review on Nov. 25, 1968, who stated 
``Adler has given an argument to the end that there exist no higher-order 
effects.  He introduced a cutoff, calculated the divergence, and then let 
the cutoff go to infinity. This is seen in the present context to be  
equivalent to the second method above.  
However, we believe that this method may not 
be reliable because of the dependence on the order of limits.''  And in 
their conclusion, they stated  ``In a definite model the nature of the 
modification (to the axial-vector current divergence equation) can be  
determined, but in general only to lowest order in interactions''. 
This controversy with 
Jackiw and Johnson was the motivation for a more thorough analysis of the 
nonrenormalization issue undertaken by Bill Bardeen and myself in the fall 
and winter of 1968--1969 (Adler and 
Bardeen [\refcite{adler-bardeen69}] ) and was cited in the ``Acknowledgments'' section of our 
paper, where we thanked ``R. Jackiw and K. Johnson for a stimulating 
controversy which led to the writing of this paper.''

The paper with Bardeen approached the problem of nonrenormalization by two 
different methods.  We first gave a general constructive argument for 
nonrenormalization of the anomaly to all orders, in both quantum 
electrodynamics and in the $\sigma$-model in which PCAC is canonically 
realized, and we then backed this argument up with an explicit calculation 
of the leading order radiative corrections to the anomaly, showing that   
they cancelled among the various contributing Feynman diagrams.  The 
strategy of the general argument was to note that since Eqs.~(\ref{aid}) and (\ref{dip}) 
involve unrenormalized fields, masses, and coupling constants, these 
equations are well defined only in a cutoff field theory.  Thus, for both 
electrodynamics and the $\sigma$-model, we constructed cutoff versions by 
introducing photon or $\sigma$-meson regulator fields with mass $\Lambda$.  
(This was simple for the case of electrodynamics, but more difficult,  
relying heavily on Bill Bardeen's prior experience 
with meson field theories, in the case of the 
$\sigma$-model.) 
In both cases, the cutoff prescription allows the usual renormalization 
program to be carried out,  expressing the  unrenormalized quantities 
in terms of renormalized ones and the cutoff $\Lambda$.  In the cutoff 
theories, the fermion loop-wise argument 
I used in my original anomaly paper is 
still valid, because regulating boson propagators does not alter the  
chiral symmetry properties of the theory, and thus it is 
straightforward to prove the validity of Eqs.~(\ref{aid}) and (\ref{dip}) for the 
unrenormalized quantities to all orders of perturbation theory.  

Taking 
the vacuum to two $\gamma$ matrix element of the anomaly equations, and 
applying the Sutherland--Veltman theorem, which asserts the vanishing 
of the matrix element 
of $\partial^{\mu}j_{\mu}^5$ at the special kinematic point $q^2=0$, 
Bardeen and I then got exact low energy theorems for the matrix elements 
$\langle 2\gamma|2im_0j^5|0\rangle$ (in electrodynamics) 
and $\langle 2\gamma|(f_{\pi}/\sqrt{2}) 
\phi_{\pi}|0 \rangle$ (in the $\sigma$-model) 
of the ``naive'' axial-vector divergence 
at this kinematic point, which were given by the negative of the 
corresponding matrix element of the anomaly term.    
However, since we could 
prove that these matrix elements are finite in the limit 
as the cutoff $\Lambda$ 
approaches infinity, this in turn gave exact low energy theorems for the 
renormalized, physical matrix elements in both cases.  One subtlety that 
entered into the all orders calculation was the role of photon rescattering 
diagrams connected to the anomaly term,  but using gauge invariance 
arguments analogous to those involved in the Sutherland--Veltman theorem, 
we were able to show that these diagrams made a vanishing contribution to 
the low energy theorem at the special kinematic point $q^2=0$.  Thus, 
my paper with Bardeen provided a rigorous underpinning for the use of the 
$\pi^0 \to \gamma \gamma$ low energy theorem to study the charge structure 
of quarks.  

In our explicit second order calculation, we calculated the leading order 
radiative corrections to this low energy theorem, arising from addition  
of a single virtual photon or virtual $\sigma$-meson to the lowest order 
diagram.  We did this by two methods, one involving a direct calculation 
of the integrals, and the other (devised by Bill Bardeen) using a clever  
integration by parts argument to bypass the direct calculation.  Both 
methods gave the same answer:  the sum of all the radiative corrections 
is zero, as expected from our general nonrenormalization argument.  We also 
traced the contradictory results obtained in the paper of Jackiw and Johnson 
to the fact that these authors had studied an axial-vector current  
(such as $\overline \psi \gamma_{\mu} \gamma_5 \psi$ in the $\sigma$-model) 
that is not made finite by the usual renormalizations in the absence 
of electromagnetism; as a consequence, the naive divergence of this current  
is not multiplicatively renormalizable.  As we noted in our paper, 
``In other words, the axial-vector current considered by Jackiw and Johnson 
and its naive divergence are not well-defined objects in the usual 
renormalized perturbation theory; hence the ambiguous results which these 
authors have obtained are not too surprising.''  Our result of a definite, 
unrenormalized low energy theorem, we noted, came about because ``In each 
model we have studied a {\it particular} axial-vector current:  in spinor 
electrodynamics, the usual axial-vector current ... and in the $\sigma$ 
model the Polkinghorne [\refcite{polk58a,polk58b}] axial-vector current ... which, in the absence 
of electromagnetism, obeys the PCAC condition.''  It is these axial-vector 
currents that obey a simple anomaly equation to all orders in perturbation 
theory, and which give an exact, physically relevant low energy theorem 
for the naive axial-vector divergence.  

This paper with Bill Bardeen should have ended the controversy over whether 
the anomaly was renormalized, but it didn't.  Johnson pointed out in an 
unpublished report that since the anomaly is mass-independent, it should 
be possible to calculate it in massless electrodynamics, for which the naive 
divergence $2im_0j^5$ vanishes and the divergence of the axial-vector 
current directly gives the anomaly.  Moreover, in massless electrodynamics 
there is no need for mass renormalization, and so if one chooses Landau gauge 
for the virtual photon propagator, the second order radiative correction 
calculation becomes entirely ultraviolet finite, with no renormalization 
counter terms needed.  Such a second order calculation was reported by 
Sen [\refcite{sen70}], a Johnson student, who claimed to find nonvanishing second 
order radiative corrections to the anomaly.   However, the calculational 
scheme proposed by Johnson and used by Sen has the problem that, 
while ultraviolet finite, there are severe 
infrared divergences, which if not handled carefully can lead to spurious 
results.  After a long and arduous calculation (Adler, Brown, Wong, and 
Young [\refcite{abwy}]) my collaborators and I were able to show that the zero mass 
calculation, when properly done, also gives a vanishing second order 
radiative correction to the anomaly.  This  confirmed the result I had found    
with Bardeen, which had by then also been confirmed by different methods 
in the $m_0\not =0$ 
theory in papers of Abers, Dicus, and Teplitz [\refcite{adt}] and Young, Wong, 
Gounaris, and Brown [\refcite{ywgb}].  

A second challenge to the nonrenormalization calculation that Bardeen and 
I had done came from DeRaad, Milton, and Tsai [\refcite{raad72}],
a group associated with  Julian Schwinger at UCLA  
\big(preprint received by Physical 
Review March 27, 1972; revised version  received by 
Physical Review May 1, 1972, and a second paper 
(Milton, Tsai, and DeRaad [\refcite{mtsai}]) with further 
calculational details written shortly afterwards\big).   
They calculated the 
radiative corrected low energy theorem using a source-theoretic method, 
and in their preprint,  
they claimed a renormalization factor of $1+\alpha/(2 \pi)$,  
in disagreement with what Bardeen and I, and other groups, had found.  
I learned of their work from Bing-Lin Young  (one of 
my collaborators in the $m_0=0$ calculation), who wrote 
to me on March 29, 1972, 
asking me to look at the UCLA group's preprint, a copy of which he enclosed.  
I did a short calculation, and then in conversations with the UCLA group 
pointed out that they had imposed an {\it ad hoc}   
normalization condition on the pseudoscalar vertex $\Gamma_5$ 
associated with the naive 
divergence $2im_0j^5$. However,    
the anomalous Ward identity for the axial-vector 
vertex part uniquely specifies the normalization to be used, since at 
$q=0$ in momentum space both the $\partial^{\mu}j_{\mu}^5$ terms and the 
anomaly term vanish, and one is left with 
$0=2m_0\Gamma_5(p,p)+S_F^{\prime}(p)^{-1} \gamma_5 
+ \gamma_5 S_F^{\prime}(p)^{-1}$. This equation relates the normalization of 
$\Gamma_5(p,p)$ to that of the electron propagator $S_F^{\prime}(p)$, and   
must be used as the normalization of the pseudoscalar vertex in computing 
radiative corrections to the low energy theorem  
\big(see equations (26) through 
(29) of my paper with Bardeen [\refcite{adler-bardeen69}]\big), even though it gives an infrared singular   
value for the on-shell vertex part $\Gamma_5(p,p)$.  DeRaad, Milton, and 
Tsai in the revised, published version of their paper [\refcite{raad72}], 
which acknowledged   
conversations with me, recalculated their  
results for a general $\Gamma_5$ normalization point and found a radiative 
correction $1+(1+\delta) \alpha/(2 \pi)$, with $\delta$ a parameter related   
to the normalization point.  They showed that the normalization 
that Bardeen and I had used corresponded to  $\delta=-1$, that is, no 
radiative correction, in agreement with our calculational result, whereas 
their original 
choice of normalization at zero momentum transfer squared 
corresponded to $\delta=0$.  They then argued 
that since the normalization that Bardeen and I had used corresponds to 
normalization at an infrared sensitive four momentum transfer $3 m^2 /
\log(\mu^2/m^2)$, with $\mu$ the fictitious photon mass, their normalization 
choice of zero four momentum transfer squared is more natural.  
However, this argument is spurious because the infrared sensitivity on which 
it is based is an artifact of the calculational method that they used.   
In fact, the vacuum to two photon 
matrix element of the naive divergence is completely infrared finite, since 
any sensitive dependence on the fictitious photon mass 
used to calculate vertex parts 
and electron propagators cancels around the closed electron loop.  Hence 
there is no justification for using a normalization convention differing from 
the one explicitly specified by the axial-vector Ward identity.  Of course, 
if one insists on using a different normalization from that determined by 
the Ward identity, a non-vanishing radiative correction to 
the anomaly will result.  As far as I can tell, Schwinger never accepted the 
nonrenormalization of the anomaly \big(see the Schwinger biography by  Mehra and Milton 
[\refcite{meh-mil}], footnote on page 488, and pp. 298-310 of Schwinger [\refcite{sch89}] \big).    

\begin{figure}[th]		
\centerline{\psfig{file=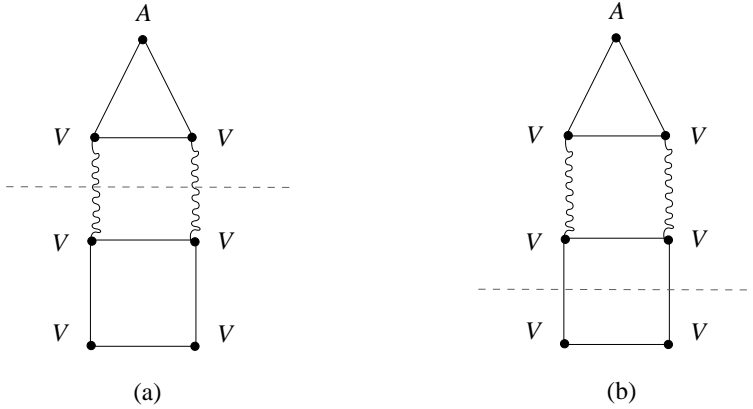,width=4in}}
\vspace*{8pt}
\caption{Photon rescattering contribution to the basic $AVV$ triangle,     
shown in (a), is identical to the triangle contribution to the axial-vector 
vertex part from Fig. 1(a) used in place of the lowest order 
axial-vector vertex in a basic $AVV$ triangle, as shown in (b).\label{fig2}}
\end{figure}
Almost two decades later, in 1989, another challenge to the Adler-Bardeen 
theorem appeared, this time in a paper of Ansel'm and Iogansen (Johansen) 
[\refcite{iogan}], again in the context of massless electrodynamics.  These authors 
showed that when the $AVV$ triangle is coupled to a light-by-light 
scattering diagram, one finds that the matrix element of $\partial^{\mu} 
j_{\mu}^5$ in a background electromagnetic field is divergent, 
\begin{equation}
\langle \partial^{\mu}j_{\mu}^5 \rangle =
\langle F_{\mu\nu}\hat F^{\mu\nu} \rangle_{ext}
(1-{3 e_0^4\over 64 \pi^4} \log{\Lambda^2 \over k^2})~~~,
\label{fan}
\end{equation}
with $k$ a typical external momentum.  This result is just what 
one would expect from using the axial-vector vertex part constructed 
from the elementary $AVV$ triangle as the axial vertex in an $AVV$ 
triangle, as shown in Fig. 2, which corresponds to the calculation done 
by Ansel'm and Iogensen.  And in fact, as noted by them, the 
divergence appearing in Eq.~(\ref{fan}) corresponds to the anomaly-induced divergence 
in the axial-vector vertex part, which is made finite by the renormalization 
factor of Eq.~(\ref{bit}) that I gave in my 1969 paper [\refcite{adler69a}], in other words, 
\begin{equation}
[1+{3\over 4}(\alpha_0/\pi)^2 \log(\Lambda^2/m^2)+...]^{-1}
=1-{3 e_0^4\over 64 \pi^4} \log{\Lambda^2/m^2}~~~.
\label{get}
\end{equation}
Moreover, Ansel'm and Iogensen stated in their paper [\refcite{iogan}] 
that the operator coefficient of 
the anomaly term has no renormalizations.  However, in asserting that the 
Adler-Bardeen theorem is incorrect, and in particular that the 
$\pi^0 \to \gamma \gamma$ low energy theorem needs radiative corrections, 
they made the 
mistake of confusing matrix elements of the axial-vector divergence 
$\partial^{\mu} j_{\mu}^5$ with matrix elements of the naive divergence 
$2im_0  j^5$.  In terms of matrix elements, the  
Adler-Bardeen theorem asserts only that the vacuum to 
two-photon matrix element of the {\it naive divergence} has a 
known value, with 
no renormalizations to all orders, at a particular kinematic point at which  
external momenta are small compared to the fermion mass.  (This kinematic  
point is needed, we recall, to be able to invoke the Sutherland-Veltman 
theorem to assert vanishing of the axial-vector divergence or $\partial^{\mu} 
j_{\mu}^5$ side of the 
anomaly equation, and to assert 
the vanishing of photon rescattering contributions from the anomaly, so 
that the naive divergence matrix element is given by the lowest order 
matrix element of {\it minus} the anomaly.)  When 
the naive divergence is the pion field, as in the anomaly-modified PCAC 
equation, the  
resulting low energy theorem has 
useful physical consequences, as discussed above.  In massless 
electrodynamics, the naive divergence operator vanishes, and the kinematic   
point at which the low energy theorem holds cannot be attained ($q^2/m^2$ is 
infinite no matter how small $q^2$ is), so there is no obvious analog of the   
$\pi^0 \to \gamma \gamma$ low energy theorem, which is probably why   
Ansel'm and Iogensen instead discuss matrix elements of 
$\partial^{\mu} j_{\mu}^5$, which in the massless case is given by  
just the anomaly term.  
However, Bardeen  and I  
{\it never} claimed that generic matrix elements of the axial-vector  
divergence,  
or the anomaly term, have no radiative corrections; such statements would  
obviously be false, 
as already shown in my 1969 anomaly paper [\refcite{adler69a}]. Thus, 
the main calculational results of the Anselm-Iogensen paper are correct, 
but they make 
interpretational statements that are incorrect, by misunderstanding  
what the Adler-Bardeen theorem says.  I gave an analysis similar to this 
in a letter 
dated July 27, 1989 to Ansel'm, and again 
in a letter dated October 13, 1997 
to Dan Freedman who passed it on to Johansen (who was visiting MIT at the 
time), but unfortunately I never received a response from either Ansel'm 
or from Johansen.  

Long before these events, however, the analysis of anomaly renormalization 
by calculation of higher order Feynman diagrams had been rendered obsolete  
by new, non-perturbative methods.   Zee [\refcite{zee72}]
and Lowenstein and Schroer [\refcite{low-sch}]  gave direct proofs of chiral 
anomaly non-renormalization in QED starting 
from the then recently discovered 
Callan-Symanzik equations, and an 
alternative argument was given in the book    
of Collins [\refcite{collins84}].  \big(For extensions to the $\sigma$-model, see also 
Becchi [\refcite{becchi}] and Shei and Zee [\refcite{szee73}].\big)  
Before describing Zee's argument,  
let me make some technical comments about the three proofs.  In the paper 
of Zee the operator $\partial A$ is to be understood as the ``naive''   
divergence of the axial-vector current;  in   
the paper of Schroer and Lowenstein this identification is made explicit.  
(When Zee's argument is applied to prove nonrenormalization 
of a flavor non-singlet current to 
all orders in strong interactions, this clarification is not needed, since 
the only anomaly term in the axial-vector divergence then 
is electromagnetic, and  
can be neglected in a strong interaction calculation.  However, when Zee's 
argument is applied to prove anomaly nonrenormalization 
in QED, the specification that $\partial A$ is just the 
naive divergence is needed.)  
The papers of Zee and of Schroer and Lowenstein both work at the 
point $q^2=0$ where the naive divergence is related to the anomaly    
coefficient.  Collins, by contrast, gives an argument based on   
renormalization properties of the axial-vector vertex holding at a 
general value of $q$.\footnote{I wish to thank Tony Zee and John Collins for phone conversations and emails on these issues.}
In his presentation, Collins 
specializes to flavor non-singlet 
axial-vector currents in QCD, such as the flavor $SU(3)$ 
octet current ${\cal F}_{3\mu}^5$ of Eq.~(\ref{dip}), 
which permits 
him to ignore complications arising from the  
gluon rescattering contributions considered by 
by Ansel'm and Iogensen. However, the Collins version of Zee's proof can 
be extended to singlet currents, such as the axial-vector current in QED or 
the flavor singlet $U(1)$ axial-vector current in QCD,  
by replacing 
the general axial-vector vertex in his argument by the {\it two photon  
irreducible}  axial-vector vertex (or its analog in QCD).  
(The general $AVV$ three-point  
function, without a two-photon irreducible restriction, is obtained 
from this simpler three-point function  by summing over the possibilities 
of zero, one, two, ... photon-photon rescatterings as shown in Fig. 3.)
The anomaly associated with the two-photon irreducible $AVV$ three-point 
function directly corresponds to the {\it operator} anomaly, since diagrams  
that iterate the two-photon irreducible three-point function with   
photon rescatterings,  as in Fig. 3,  are in one-to-one 
correspondence with diagrams that iterate the operator  
anomaly with photon rescatterings.  


\begin{figure}[th]		
\centerline{\psfig{file=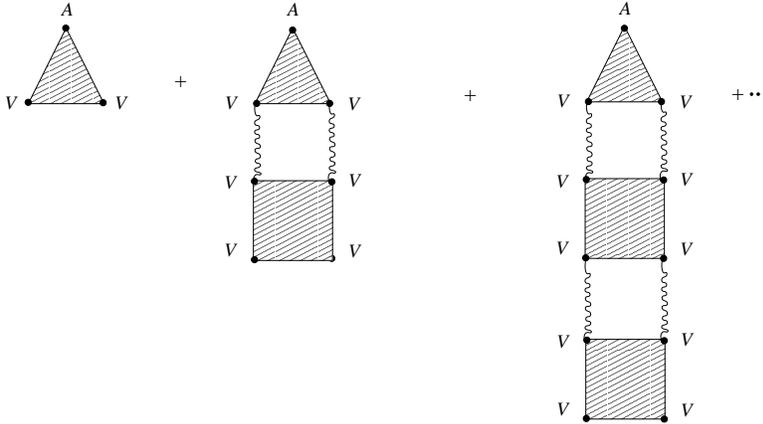,width=4in}}
\vspace*{8pt}
\caption{The full $AVV$ three-point function is obtained from the   
two-photon irreducible $AVV$ three-point function (shaded triangle) 
by adding the contributions 
of one or more photon rescatterings.   \label{fig3}}
\end{figure}

Briefly, Zee's argument is based on consideration of the three-point 
function $m_0j^5V_{\mu}V_{\nu}$ formed from the naive axial current 
divergence and from 
vector currents carrying four-momenta $k_1$, $k_2$, which he denotes in   
momentum space by 
$R_{D\mu\nu}(k_1,k_2)$. From a combination of the Callan-Symanzik equation  
for this three-point function, and the Ward identity for the four-point 
function obtained by inserting a scalar 
current in this three-point function, Zee establishes the 
formula
\begin{equation}
\left( m{\partial \over \partial m} 
+ \alpha \beta(\alpha){\partial\over \alpha}\right)
R_{D\mu\nu}(k_1,k_2) \propto O\big( (k_1+k_2)k_1k_2\big)~~~.
\label{hip}
\end{equation}
Hence expanding in powers of $k_1$ and $k_2$, by writing  
$R_{D\mu\nu}(k_1,k_2)=f\epsilon_{\mu\nu\rho\sigma}k_1^{\rho}k_2^{\sigma}
+...$, the coefficient $f$ of the leading term obeys 
\begin{equation}
\left( m{\partial \over \partial m} 
+ \alpha \beta(\alpha){\partial\over \alpha}\right) f=0~~~.
\label{ice}
\end{equation}
Since the dimensionless coefficient $f$ must be independent of $m$, 
this implies that $\partial f/\partial 
\alpha=0$, and so $f$ has no dependence on $\alpha$!  As Zee notes in 
his paper, this argument 
extends readily to theories in which there is more than one nonvanishing 
mass, since there are then additional renormalization group equations to   
give constraints.  
Thus, the renormalization group   
argument shows that the low energy theorem for the naive divergence, and 
correspondingly the coefficient of the operator anomaly, 
is given by the lowest order 
triangle contribution, and extends the Adler-Bardeen theorem to rule out 
even non-perturbative renormalizations of the anomaly.  

Recent textbooks often present the anomaly through the elegant path 
integral formulation given by Fujikawa [\refcite{fujik79,fujik80a}].  In this formulation,  
as applied to QED, 
one starts from a Feynman path integral for the partition function 
\begin{equation}
Z=\int d[B_{\mu}] d[\psi] d[\overline \psi]
e^{-S[B,\psi,\overline\psi]}~~~,
\label{jam}
\end{equation}
with $B_{\mu}$ the electromagnetic potential four-vector and with $\psi$ 
the fermion field. One then makes a chiral 
transformation on the fermion fields, for which the axial-vector current 
is the Noether current, and examines the behavior of the   
various terms in the path integral.  Since all variables in the functional 
integral are classical, the change in the mass term in the 
action $S$ yields just the naive divergence of the axial-vector current.  
Fujikawa's crucial observation is that this is not the end of the story, 
because one also has to calculate the Jacobian associated with the chiral 
transformation of the functional integration measure.  Since this measure 
is an infinite product, a regularization is needed to define it, and as a 
result of the regularization, the measure is not a chiral invariant.  
Within 
a suitable family of regularizations, Fujikawa shows that the Jacobian 
associated with the chiral transformation just produces an extra phase 
factor corresponding to the axial-vector anomaly! One can now ask the  
question, does Fujikawa's method provide a simple and direct 
non-renormalization proof, or is the fact that his calculation gives the 
exact answer a {\it consequence} of the fact that the lowest order anomaly 
is exact?  This question has been examined by Shizuya and Tsukahara [\refcite{shiz-tsuk}], 
Shizuya [\refcite{shiz87}], 
and Joglekar and Misra [\refcite{jog-misra}], who show that Fujikawa's result is a 
one-loop result, and that additional input from perturbation theory and/or 
details of the regulator structure is needed 
to show that the higher order corrections to the anomaly, arising formally 
in the path integral method, all vanish. This conclusion is reinforced by   
the example of the trace anomaly, to be discussed in Section~\ref{trace}, where there 
are higher order (but known) perturbative corrections to the anomaly, with  
the path integral Jacobian argument giving only the leading order term.  
Brief discussions of the relation between the path integral method and the 
calculation of anomalies to all orders are also found  
in a Festschrift article by Fujikawa [\refcite{fujik87}], and in a recent paper 
of Arkani-Hamed and Murayama [\refcite{ahm}] (see also hep-th/9707133).

To conclude this section on the quantum electrodynamics or 
Abelian anomaly, we mention a recent 
interesting paper by Vainshtein [\refcite{va03}], and its extension and 
application to the muon $g-2$ calculation 
by Czarnecki, Marciano, and Vainshtein [\refcite{czar03}] (see also hep-ph/0310276), and by Knecht,
Perrottet, de Rafael, and Peris [\refcite{kpdp04}] (for relevant earlier work by these authors, see
[\refcite{kpdp02}]).  Vainshtein considers 
the two-photon irreducible triangle 
$A_{\mu}V_{\sigma}V_{\tau}$  with one photon soft, and 
defines an amplitude 
$T_{\mu \sigma}$ obtained from this triangle by contracting it with  
the soft photon polarization $e^{\tau}$.  This two-index tensor amplitude 
can be divided into longitudinal and transverse parts.  In the chiral 
(zero fermion mass) limit, there is no naive divergence and so the 
longitudinal part is given by the anomaly, and is known to all orders. 
Vainshtein points out that in the chiral limit, there is a crossing symmetry 
relating the vector and axial-vector indices $\mu$ and $\sigma$, since 
the $\gamma_5$ in the axial vertex with index $\mu$ 
can be freely anticommuted through 
fermion propagators and vector vertices 
until it is part of the external vector vertex with index $\sigma$.  Applying  
this crossing symmetry to the absorptive part of $T_{\mu \sigma}$,  
together with a dispersion relation argument, implies that the transverse 
part $w_T$ of $T_{\mu \sigma}$ is equal to one half of the longitudinal 
part $w_L$, and thus also has no perturbative radiative corrections in 
higher orders.  As discussed in the second paper cited above, this result  
has applications to the calculation of hadronic corrections to the muon 
$g-2$, in which a triangle diagram is joined to the muon line by a virtual 
photon and a virtual $Z$ intermediate boson.  

\subsection{Point Splitting Calculations of the Anomaly. Did Schwinger Discover the Anomaly? \label{point-split}}

At this point let me backtrack, and discuss the role of point-splitting 
methods in the study of the Abelian electrodynamics anomaly.  In the present 
context, point-splitting was first used in the discussion 
given by Schwinger [\refcite{sch51}] of the
pseudoscalar-pseudovector equivalence theorem, to be described in more 
detail shortly.  Almost immediately 
following circulation of the seminal anomaly preprints 
in the fall of 1968, Hagen [\refcite{hagen69}]( received Sept. 24, 1968, and a
letter to me dated Oct. 16, 1968), Zumino [\refcite{zumino69}] 
(and a letter to me dated Oct. 7, 1968), and Brandt [\refcite{brandt}](received  
Dec. 17, 1968, and a letter to me dated Oct. 16, 1968) 
all rederived the anomaly formula by a point-splitting method.  
Independently,  a point-splitting  
derivation of the anomaly was given by Jackiw and 
Johnson [\refcite{jackjohn}] (received 25 November, 1968), who explicitly made the   
connection to Schwinger's earlier work (Johnson was a Schwinger student, and 
was well acquainted with Schwinger's body of work).  
The point of all of 
these calculations is that the anomaly   
can be derived by formal algebraic use of the equations of motion, 
provided one redefines 
the singular product $\overline \psi(x) \gamma_{\mu} \gamma_5 \psi(x)$ 
appearing in the axial-vector current by the point-split expression 
\begin{equation}
\lim_{x \to x^{\prime}} \overline \psi(x^{\prime}) \gamma_{\mu} \gamma_5 
\exp[-ie\int_{x^{\prime}}^x dx^{\lambda} B_{\lambda}] \psi(x)~~~,
\label{kit}
\end{equation}
and takes the limit $x^{\prime} \to x$ at the end of the calculation. 

Responding to these developments, 
I appended a ``Note added in proof'' to my anomaly paper, mentioning the 
four  field-theoretic, point-splitting derivations 
that had subsequently been 
given of Eq.~(\ref{aid}), and adding  ``Jackiw and Johnson point out that the 
essential features of the field-theoretic derivation, in the case of 
external electromagnetic fields, are contained in J. Schwinger, Phys. 
Rev. {\bf 82}, 664 (1951)''.  
What to me was an interesting irony emerged from learning  
of the connection between anomalies and the 
famous Schwinger paper [\refcite{sch51}] on vacuum polarization.    
I had in fact read Section II 
and the Appendices of the 1951 paper, when Alfred Goldhaber and I, during  
our senior year at Harvard (1960-61), 
did a reading course on quantum electrodynamics 
with Paul Martin, which focused on papers in Schwinger's reprint volume (Schwinger [\refcite{sch58}]).  Paul had told us to read the parts of the Schwinger paper 
that were needed to calculate the $VV$ vacuum polarization loop, but 
to skip the rest as being too technical.  Reading Section V of Schwinger's 
paper brought back to mind a brief, forgotten conversation I had had with 
Jack Steinberger, who was Director of the Varenna Summer School in 1967.  
Steinberger had told me that he had done a calculation 
on the pseudovector-pseudoscalar 
equivalence theorem for $\pi^0 \to \gamma \gamma$, but had 
gotten different answers in the two cases; also that Schwinger had claimed 
to reconcile the answers, but that he (Steinberger) couldn't make sense out 
of Schwinger's argument.  Jack had urged me to look at it, which I never did 
until getting the Jackiw--Johnson preprint, but in retrospect everything 
fell into place, and the connection to Schwinger's work was apparent.    

This now brings me to the question, did Schwinger's paper constitute  
the discovery of  
the anomaly?  Both Jackiw, in his paper with Johnson, and I were 
careful to note the 
connection between Schwinger's paper [\refcite{sch51}]  and the point-splitting 
derivations of the anomaly, once it was called to our attention.  However, 
recently a number of Schwinger's former students have gone further, arguing 
that Schwinger was the  discoverer of the anomaly and that my paper and 
that of Bell and Jackiw were merely a ``rediscovery'' of a previously 
known result. For  example, 
 Mehra and Milton [\refcite{meh-mil}] refer (p. 488) to ``...the 
axial-vector anomaly$^*$ which he (Schwinger) had discovered in [64] 
(their reference number for the 1951 paper)'', and in the footnote labeled by $^*$ on the same page say ``As we have 
noted, Stephen Adler, John S. Bell, and Roman Jackiw had rediscoverd the 
anomaly in 1968 ''.   In a similar vein, 
when I visited the University of Washington a few years ago I noticed a 
display case that had been set up in the theory wing, featuring the
1951 paper and arguing that Schwinger was the discoverer of the anomaly.  

With due respect to these opinions of colleagues, I disagree, for  
a number of reasons.  The essence of the anomaly is that {\it quantization 
necessarily violates classical symmetries}, or to borrow a phrase from  
Bjorken's referee's report quoted above, not every ``cherished symmetry 
of the theory'' can be maintained.  For example, in quantum theory 
the pseudovector 
coupling calculation of $\pi^0 \to 2\gamma$ decay, which 
by the Sutherland-Veltman theorem 
gives zero, and the pseudoscalar coupling calculation, which 
gives nonzero, do  not agree, even though 
classically there is a pseudoscalar-pseudovector equivalence 
theorem.
This disagreement arises, as discussed in detail 
above, because the naive axial divergence equation used in proving the 
equivalence theorem is invalid in the quantized theory, as manifested   
in the anomaly.  
By contrast, Schwinger's 
calculation was devoted to making 
the pseudovector calculation give {\it the same} non-zero answer as the 
pseudoscalar one.\footnote{I wish to thank Peter van   
Nieuwenhuizen for emphasizing this point, as well as for stressing that 
Schwinger's calculation is not fully gauge invariant, and for urging me to 
make precise the relation between it and the later point-splitting 
calculations of the anomaly.}
He did this by redefining the divergence of the 
axial-vector current as the limit of a point-split, gauge-covariant 
divergence (using his notation here, except that we continue to denote the 
electromagnetic vector potential by $B_{\mu}$), 
\begin{equation}
\partial_{\mu}[{\rm tr} \gamma_5 \gamma_{\mu}G(x,x)] 
\equiv\lim_{x^{\prime},x^{\prime\prime} \to x}
[(\partial_{\mu}^{\prime}-ieB_{\mu}(x^{\prime})) 
+(\partial_{\mu}^{\prime\prime}+ieB_{\mu}(x^{\prime\prime}))] 
{\rm tr} \gamma_5 \gamma_{\mu}G(x^{\prime},x^{\prime\prime}) ~~~,
\label{lad}
\end{equation}
with $G(x^{\prime},x^{\prime\prime})$ the electron propagator. As noted 
in  Appendix A, this equation is simply a rewriting of the pseudoscalar 
coupling expression $-2m{\rm tr} \gamma_5G(x^{\prime},x^{\prime\prime})$ 
by use of the equations of motion.  Thus Schwinger's calculation is an 
indirect way of calculating the effective Lagrangian for pseudoscalar 
coupling, in the process of which he derives identities that 
can also be used to give a point-splitting calculation of the anomaly, 
as discussed, in the context of Zumino's calculation, in Appendix A.  
Although 
the expression of Eq.~(\ref{lad}) is only gauge covariant, it can 
be made gauge invariant 
by multiplying by an overall Wilson line factor 
$\exp[-ie\int_{x^{\prime\prime}}^{x^{\prime}} dx^{\lambda} B_{\lambda}]$ as 
in Eq.~(\ref{kit}), which since it is outside the derivatives reduces to unity 
in the limit as $x^{\prime\prime}\to x^{\prime}$.  With the overall  
Wilson line factor included, Schwinger's formulas are all gauge invariant, 
and one sees that what Schwinger calls the redefined axial-vector 
divergence is in fact, when expressed in the notation used in Section~\ref{stage},  
{\it not} the divergence of the gauge-invariant 
axial vector current, but rather the axial-vector current  
divergence {\it minus} the anomaly.  
In other words, Schwinger's calculation effectively transposes the 
anomaly term to the left hand side of Eq.~(\ref{aid}), so that what he  
evaluates is the effective Lagrangian arising from the left hand side of 
the equation 
\begin{equation}
\partial^{\mu} j_{\mu}^5(x) 
- {\alpha_0 \over 4 \pi} 
F^{\xi \sigma}(x) F^{\tau \rho}(x) \epsilon_{\xi\sigma \tau \rho}
= 2im_0j^5(x) ~~~,
\label{met}
\end{equation}
which then gives the same result as calculation of an effective Lagrangian 
from the right hand side, which is pseudoscalar coupling.  

The use of a point-splitting method was of course 
important and fruitful, and in retrospect, the axial anomaly is hidden 
within Schwinger's calculation.  But Schwinger never 
took the crucial step of observing that the axial-vector current matrix 
elements cannot, in a renormalizable quantum theory, be made to satisfy 
all of the expected classical symmetries.  And more specifically, he   
never took the step of defining a gauge-invariant axial-vector 
current by point splitting, as in Eq.~(\ref{kit}), 
which has a well-defined anomaly term in  
its divergence, as in Eq.~(\ref{aid}), with the anomaly term  completely 
accounting for the disagreement 
between the pseudoscalar and pseudovector calculations of neutral pion 
decay.  So I would say that although Schwinger took steps in the right 
direction, particularly in noting the utility of point-splitting in 
defining the axial-vector current, his 1951 paper {\it obscured} the 
true physics and does  not mark the discovery of the anomaly. 
This happened only much later, in 1968, and led to a flurry of activity 
by many people.  My view is supported, I believe, by the 
fact that Schwinger's calculation seemed arcane, even to 
people (like Steinberger)  with whom he had talked about it  
and to colleagues familiar with his work,  
and exerted no influence on the field until after preprints on the seminal   
work of 1968 had appeared.  

\subsection{The Non-Abelian Anomaly, its Nonrenormalization and Geometric Interpretation \label{naa}}

Since in the chiral limit the $AVV$ triangle is identical to an $AAA$ 
triangle (by a similar argument to that explained at the end of Section~\ref{anom-non}, 
involving anticommutation of a $\gamma_5$ 
around the loop), I knew 
already in unpublished notes dating from the late summer of 1968 that 
the $AAA$ triangle would also have an anomaly; a similar observation was   
also made by Gerstein and Jackiw [\refcite{gerjack}].  
>From fragmentary calculations begun in 
Aspen I suspected that higher loop diagrams might have anomalies as well, 
so after the nonrenormalization work with Bill Bardeen was finished I  
suggested to Bill that he work out the general anomaly for larger 
diagrams. (I was at that point involved in other calculations with  
Wu-ki Tung, on the perturbative breakdown of scaling formulas such as 
the Callan-Gross relation.)  I showed Bill my notes, which turned out to 
be of little use, but which 
contained a very pertinent remark by Roger Dashen that including charge 
structure (which I had not) would allow a larger class of potentially 
anomalous diagrams.   Within a few weeks, Bill carried out a 
brilliant calculation, by point-splitting methods, of the general anomaly 
in both the Abelian {\it and} the non-Abelian cases (Bardeen [\refcite{bar69}]).  

The work discussed in Section~\ref{point-split} on the use of point-splitting in the 
Abelian case was known to 
Bardeen when he started his calculation, and influenced his choice of 
methodology.  Rather than proceeding directly from Feynman diagrams as in  
my anomaly paper and 
that of Bell and Jackiw, he 
proceeded by combining a generalized version of point splitting with   
the formal $S$-matrix expansion.  (For a subsequent 
textbook derivation of the general 
anomaly directly from Feynman diagrams, see Chapt. 1 of 
van Nieuwenhuizen [\refcite{vnp88}].)
Bardeen's ingenious adaptation of point-splitting involved using 
different infinitesimal separations for loops with different numbers 
of vertices. Thus, he started from the  
interaction Lagrangian ${\cal L}(z) = 
\overline \psi(z) \Gamma(z) \psi(z)$, with $\Gamma(z)$ involving general
nonderivative couplings to external scalar, pseudoscalar, vector, and 
axial-vector fields, and replaced it with 
\begin{equation}
{\cal L}^{\epsilon}(z)=\overline \psi \left( z+{\epsilon\over 2n}\right) 
\Gamma(z) \psi \left( z-{\epsilon\over 2n}\right) ~~~,
\label{nap}
\end{equation}
with $n$ the number of vertices in the loop, and with an average over 
$\epsilon$ to be performed after evaluation of the loop.  As 
explained by Bardeen, ``The use of $1/n$ in the definition of the $\epsilon$ 
separation is necessary so that loops with different numbers of vertices 
may be simply related, as needed for the Ward identities.''

Bardeen then defined vector and axial-vector currents by a variation of the 
$S$ matrix with respect to the corresponding external fields, and calculated 
all of the Ward identities involving the smaller spinor loops.  These 
contained singular terms and non-minimal ``anomaly'' terms that 
were artifacts of the 
calculational method, both of which could be removed by adding local 
polynomial counterterms to the single loop $S$ matrix.  What remained after 
this redefinition were the true anomaly terms.  Again quoting from Bardeen, 
``If we required that the vector currents have the normal divergences, the 
divergences of the axial-vector currents contained the minimal anomalous 
terms given in Eq.~(49).  These anomalous terms were minimal in the sense 
that any further redefinition of the $S$ matrix  would either destroy the 
normal vector-current divergences or simply give additional terms in the 
anomalous axial-vector-current divergences.''
Expressed in terms of vector and axial-vector Yang-Mills field strengths 
\begin{eqnarray}
\label{opp}
F_V^{\mu\nu}(x)=&\partial^{\mu}V^{\nu}(x)-\partial^{\nu}V^{\mu}(x) 
-i[V^{\mu}(x),V^{\nu}(x)]-i[A^{\mu}(x),A^{\nu}(x)]~~~,\nonumber \\
F_A^{\mu\nu}(x)=&\partial^{\mu}A^{\nu}(x)-\partial^{\nu}A^{\mu}(x) 
-i[V^{\mu}(x),A^{\nu}(x)]-i[A^{\mu}(x),V^{\nu}(x)]~~~,
\end{eqnarray}
Bardeen's famous Eq.~(49) for the non-Abelian axial-vector anomaly takes 
the form 
\begin{eqnarray}
\label{par}
\lefteqn{\partial^{\mu}J_{5\mu}^{\alpha}(x)={\rm normal~divergence~term}}\nonumber \\
&&+(1/4\pi^2)\epsilon_{\mu\nu\sigma\tau}{\rm tr}_I \lambda_A^{\alpha}
[(1/4)F_V^{\mu\nu}(x)F_V^{\sigma \tau}(x)
+(1/12)F_A^{\mu\nu}(x)F_A^{\sigma \tau}(x) ~~~\nonumber \\
&&+(2/3)iA^{\mu}(x)A^{\nu}(x)F_V^{\sigma \tau}(x) 
+(2/3)i F_V^{\mu \nu}(x) A^{\sigma}(x)A^{\tau}(x) 
+(2/3)i  A^{\mu}(x)F_V^{\nu \sigma}(x) A^{\tau}(x) \nonumber \\
&&-(8/3) A^{\mu}(x)A^{\nu}(x)A^{\sigma}(x)A^{\tau}(x) ]~~~,
\end{eqnarray}
with ${\rm tr}_I$ denoting a trace over internal degrees of freedom, and   
$\lambda_A^{\alpha}$ the internal symmetry matrix associated with 
the axial-vector external field.  In the Abelian case, with trivial  
internal symmetry structure, 
the terms involving two or three factors of $A^{\mu,\nu,...}$ vanish by 
antisymmetry of $\epsilon_{\mu\nu\sigma\tau}$, and there are only 
$AVV$ and $AAA$ triangle anomalies.  When there is non-trivial internal 
symmetry or charge 
structure, there are anomalies associated with the box and pentagon 
diagrams as well, confirming Dashen's intuition mentioned earlier.  
Bardeen notes that whereas the triangle and box anomalies result from linear 
divergences associated with these diagrams, the pentagon anomalies 
arise not  from linear divergences, but rather from the definition of the 
box diagrams to have the correct vector current Ward identities.  
Bardeen also notes, in his 
conclusion, another prophetic remark of Dashen, to the effect that the 
pentagon anomalies should add anomalous terms to the PCAC low energy 
theorems for five pion scattering.  

There are two distinct lines of argument leading to the conclusion that 
the non-Abelian chiral anomaly also has a nonrenormalization theorem, and 
is given exactly by Bardeen's leading order calculation.  The first 
route parallels that used in the Abelian case, involving 
a loop-wise regulator construction in the proof initially given by Bardeen [\refcite{bar72,bar74}], followed by explicit fourth order calculation (Chanowitz [\refcite{chan74}]), an $SU(3)$ 
analog of the Zee argument using the Callan-Symanzik equations 
(Pi and Shei [\refcite{pishei}]), and use of dimensional regularization (Marinucci and Tonin [\refcite{mt}], Costa, Julve, Marinucci, and Tonin [\refcite{cjmt}]). The conclusion in all cases is that the Adler-Bardeen 
theorem extends to the non-Abelian case.  Heuristically, what is happening 
is that except for a few small one-fermion loop diagrams, 
non-Abelian theories, just like Abelian ones, are made 
finite by gauge invariant regularization of the gluon 
propagators.  But this regularization has no effect on the chiral properties 
of the theory, and therefore does not change its anomaly structure, which     
can thus be deduced from the structure of the few small fermion loop 
diagrams for which naive classical manipulations break down.    
For a later, more mathematically oriented proof, and further references,  
see Lucchesi, Piguet, and Sibold [\refcite{lucch87}].  

The second route leading to the conclusion that the non-Abelian anomaly 
is nonrenormalized might be termed ``algebraic/geometrical'', and consists 
of two steps.  The first step consists of a demonstration that 
the higher order terms in Bardeen's non-Abelian formula are completely 
determined by the leading, Abelian anomaly.  During a summer visit to  
Fermilab in 1971, I collaborated with Ben Lee, Sam Treiman, and Tony Zee [\refcite{altz71}]
in a calculation of a low energy theorem for the 
reaction $\gamma + \gamma \to \pi + \pi + \pi$ in both the neutral and charged
pion cases.  This was motivated in part by discrepancies in calculations 
that had just appeared in the literature, and in part by its relevance to 
theoretical unitarity calculations of a lower bound on the $K_L^0 \to 
\mu^+ \mu^-$ decay rate.  Using PCAC, we showed that the fact that the 
$\gamma + \gamma \to 3 \pi$ matrix elements vanish in the limit when a 
final $\pi^0$ becomes soft, together with photon gauge invariance, relates  
these amplitudes to the matrix elements $F^{\pi}$ for 
$\gamma +\gamma \to \pi^0$ and $F^{3\pi}$ for $\gamma \to \pi^0 + 
\pi^+ + \pi^-$, and moreover, gives a relation between the latter two 
matrix elements, 
\begin{equation}
eF^{3 \pi} =f^{-2} F^{\pi}~~~,~f={f_{\pi}\over 
\sqrt 2 \mu_{\pi}^2}~~~.
\label{que}
\end{equation} 
Thus all of the matrix elements in question are uniquely determined by 
$F^{\pi}$, which itself is determined by the $AVV$ anomaly calculation.  
An identical result for the same reactions was independently given by 
Terent'ev (Terentiev) [\refcite{terent71}].  In the meantime, in a 
beautiful formal analysis, 
Wess and Zumino [\refcite{wesszum}]  showed that the current algebra satisfied by the 
flavor $SU(3)$ octet of vector and axial-vector currents implies 
integrability or ``consistency'' conditions on the non-Abelian 
axial-vector anomaly, which are satisfied 
by the Bardeen formula, and conversely, that these constraints 
uniquely imply the Bardeen structure 
up to an overall factor, which is determined by the Abelian $AVV$ anomaly.   
By introducing an auxiliary pseudoscalar field, Wess and Zumino were able 
to write down a local action obeying the anomalous Ward identities and the  
consistency conditions.  (There is no corresponding local action involving 
just the vector and axial-vector currents, since if there were, the 
anomalies could be eliminated by a local counterterm.)  
Wess and Zumino also gave expressions for the processes 
$\gamma \to 3 \pi$ and $2 \gamma \to 3 \pi$ discussed by Adler et al  
and Terentiev, as well as giving the anomaly contribution to the five 
pseudoscalar vertex.  The net result of these three simultaneous pieces of 
work was to show that the Bardeen formula has a rigidly constrained 
structure, up to an overall factor given by the $\pi^0 \to \gamma \gamma$ 
decay amplitude.  

The second step in the ``algebraic/geometric'' route to anomaly 
renormalization is a celebrated paper of Witten [\refcite{ewitt83}], which shows  
that the Wess-Zumino action has a  representation as the integral of 
a fifth rank antisymmetric tensor (constructed from the auxiliary 
pseudoscalar field) 
over a five dimensional disk of which four dimensional space is the 
boundary.  In addition to giving a new interpretation of the Wess-Zumino 
action $\Gamma$, Witten's argument also gave a constraint on the 
overall factor in $\Gamma$ that was not determined by the Wess-Zumino    
consistency argument.  Witten observed 
that his construction is not unique, because a closed five-sphere 
intersecting a hyperplane gives two ways of bounding the four-sphere along 
the equator with a five dimensional hemispherical disk.  Requiring these two constructions 
to give the same value for  
$\exp(i\Gamma)$, which is the way the anomaly enters into a Feynman 
path integral, requires integer quantization of the overall coefficient 
in the Wess-Zumino-Witten action.  This integer can be read off from 
the $AVV$ triangle diagram, and for the case of an underlying color 
$SU(N_c)$ gauge theory turns out to be just $N_c$, the number of colors.     

To summarize, the ``algebraic/geometric'' approach shows that the Bardeen 
anomaly has a unique structure, up to an overall constant, and moreover that 
this overall constant is constrained by an integer 
quantization condition. Hence once the overall constant is fixed by 
comparison with leading order perturbation theory (say in QCD), it is 
clear that this  
result must be exact to all orders, since the presence of renormalizations 
in higher orders of the strong coupling constant would lead to violations 
of the quantization condition.   

The fact that non-Abelian anomalies are given by an overall rigid structure 
has important implications for quantum field theory.  For example,  
the presence of anomalies spoils the renormalizability of non-Abelian 
gauge theories and requires the cancellation
of gauged anomalies between different fermion species,\footnote{See 
Gross and Jackiw [\refcite{gross72}], Bouchiat, Iliopoulos and Meyer [\refcite{bim}], and 
Weinberg [\refcite{wein73}] .} through imposition   
of the condition ${\rm tr} \{T_{\alpha},T_{\beta}\} T_{\gamma}=0$ for all 
$\alpha,\beta,\gamma$, with $T_{\alpha}$ the coupling matrices of gauge 
bosons to left-handed fermions.  
The fact that  
anomalies have a rigid structure then implies that once these anomaly 
cancellation conditions are imposed for the lowest order anomalous triangle 
diagrams, no further conditions arise from anomalous square or pentagon  
diagrams, or from radiative corrections to these leading fermion loop
diagrams.  A second place where the one-loop geometric 
structure of non-Abelian 
anomalies enters is in instanton physics, where the structural match 
between the instanton action $8\pi^2/g^2$ and the anomaly coefficient 
$g^2/(16 \pi^2)$ results in the integral of the anomaly over an instanton 
being an integer.  This integer can be interpreted as a 
topological winding number, and through  
the anomaly leads to a relation between chiral transformations of the 
fermion fields $\psi_f \to \exp(i\gamma_5\alpha_f) \psi_f$, and shifts in 
the theta angle that describes the gauge 
field vacuum, $\theta \to \theta+2\sum_f \alpha_f$.  
The fact that the non-Abelian anomaly is nonrenormalized implies that 
this relation is exact, and has no higher order 
perturbative corrections.\footnote{I have followed here 
the notation used in the text of Weinberg [\refcite{wein96}], p. 457.  I am indebted to Hitoshi Murayama for this remark on instanton physics.}  Yet another place where anomaly nonrenormalization plays a role is in the 't Hooft anomaly matching conditions \big(for a survey and 
references, see Weinberg [\refcite{wein96}], Sec. 22.5\big).

\section{Trace Anomalies and their all Orders Calculation \label{trace}}

In an influential paper Wilson [\refcite{kgw69}] proposed the 
operator product expansion, 
incorporating ideas on the approximate scale invariance of the strong 
interactions suggested by Mack [\refcite{gmack}].  As one of the applications of 
his technique, Wilson discussed $\pi^0 \to  2\gamma$ decay and the 
axial-vector anomaly from 
the viewpoint of the short distance singularity of the coordinate space 
$AVV$ three-point function. Using these methods, Crewther [\refcite{crew72}]  and 
Chanowitz and Ellis [\refcite{chan72}] investigated the short distance structure of 
the three-point function  $\theta V_{\mu}V_{\nu}$, 
with $\theta=\theta_{\mu}^{\mu}$ 
the trace of the energy-momentum tensor, and computed the explicit form of its leading order anomaly.\footnote{In Crewther's Eqs.~(14) and (15) the trace   
$\theta_{\mu}^{\mu}$ is the ``naive'' trace excluding the electromagnetic 
anomaly contribution, that is, it only contains hadronic fields. I wish  
to thank Rod Crewther for email correspondence on this point.  
(My comment here   
is analogous to the comment above that $\partial A$ in Zee's paper [\refcite{zee72}]  
is the ``naive'' divergence.)} Their calculations confirmed earlier indications of a perturbative trace anomaly obtained in a study of broken scale invariance by Coleman and Jackiw [\refcite{coljac}]. 

Letting $\Delta_{\mu\nu}(p)$ be the momentum space 
$\theta V_{\mu}V_{\nu}$ three-point function, and $\Pi_{\mu\nu}$ be 
the corresponding $V_{\mu}V_{\nu}$ two-point function, the naive Ward 
identity $\Delta_{\mu\nu}(p)=(2-p_{\sigma}\partial/\partial p_{\sigma}) 
\Pi_{\mu\nu}(p)$ is modified to 
\begin{equation}
\Delta_{\mu\nu}(p)=\left(2-p_{\sigma}{\partial\over \partial p_{\sigma}}
\right) \Pi_{\mu\nu}(p)-{R\over 6 \pi^2}(p_{\mu}p_{\nu}-\eta_{\mu\nu}p^2) 
~~~~,
\label{ran}
\end{equation} 
with the trace anomaly coefficient $R$ given by 
\begin{equation}
R=\sum_{i,{\rm spin} {1\over 2}}Q_i^2 
+{1\over 4} \sum_{i,{\rm spin} 0}Q_i^2~~~.
\label{sis}
\end{equation}  
Thus, for QED, with a single fermion of charge $e$, 
the anomaly term is $-[2\alpha/(3 \pi)] (p_{\mu}p_{\nu}-\eta_{\mu\nu}p^2)$.
In a subsequent paper, Chanowitz and Ellis [\refcite{chan73}] 
showed that the fourth order  
trace anomaly can be read off directly from the coefficient of the leading 
logarithm in the asymptotic behavior of $\Pi_{\mu\nu}(p)$, giving to next 
order an anomaly coefficient $ -2\alpha/(3 \pi) -\alpha^2/(2 \pi^2)$.  
Thus, their fourth order argument indicated a direct connection 
between the trace anomaly and the renormalization group $\beta$ function.  

Although not the historical route, the fact that there is a trace anomaly 
can also be inferred by inspection of the Pauli-Villars regulator 
construction used to define gauge invariant fermion loop diagrams.   
As pointed out by Zumino [\refcite{zuminonew69}] and by 
Jackiw [\refcite{jack70}], the axial-vector anomaly can be given a 
simple interpretation this way.  Introducing a Pauli-Villars fermion 
of mass $M_0$, the vacuum to two photon matrix element of the 
regularized axial-vector divergence equation takes the form 
\begin{equation}
\langle 0|\partial^{\mu}j_{\mu}^5|\gamma \gamma \rangle^{m_0}   
- \langle 0|\partial^{\mu}j_{\mu}^5|\gamma \gamma \rangle^{M_0}   
=2im_0 \langle 0|j^5|\gamma \gamma\rangle^{m_0} 
- 2iM_0\langle 0|j^5|\gamma \gamma\rangle^{M_0}~~~.
\label{tap}
\end{equation}   
The left hand side of this equation approaches the regularized axial-vector 
divergence as $M_0\to \infty$.  The first term on the right hand side 
is well defined by itself and gives the matrix element of the naive 
axial-vector divergence.  
However, gauge invariance combined with 
dimensional analysis (or direct calculation) shows that for large 
$M_0$, we have $\langle 0|j^5|\gamma \gamma\rangle^{M_0}\propto 
M_0^{-1}\langle 0|F^{\xi \sigma} F^{\tau \rho} \epsilon_{\xi\sigma \tau \rho}|\gamma \gamma \rangle^{M_0}$, and so the second term on the right approaches a nonzero 
limit as $M_0 \to \infty$, and gives the anomaly contribution.  Let us 
now apply a similar analysis to the vacuum to two photon matrix element 
of the energy-momentum tensor, defined in a similar way by Pauli-Villars 
regularization, 
\begin{equation}
\langle 0|\theta_{\mu}^{\mu}|\gamma \gamma \rangle^{m_0}   
- \langle 0|\theta_{\mu}^{\mu}|\gamma \gamma \rangle^{M_0}   
=m_0 \langle 0|j|\gamma \gamma\rangle^{m_0}
- M_0\langle 0|j|\gamma \gamma\rangle^{M_0}~~~,
\label{ute}
\end{equation}  
with $j=\overline \psi \psi$ the scalar current.   With the imposition of gauge invariance (to eliminate quadratic divergences), the left hand 
side of this equation approaches the regularized trace of the energy-momentum 
tensor, and the first term on the right is well defined and 
gives the matrix element of the naive trace.  However, for the scalar   
current, gauge invariance combined with 
dimensional analysis shows that for large 
$M_0$, we have $\langle 0|j|\gamma \gamma\rangle^{M_0}\propto 
M_0^{-1}\langle 0|F^{\rho \sigma} F_{ \rho \sigma}| \gamma \gamma\rangle^{M_0}$, 
and so the second term on the right approaches a nonzero 
limit as $M_0 \to \infty$, and gives the trace anomaly. Pursuing the  
details by computing the proportionality constants, one sees that Eqs.~(\ref{tap},\ref{ute}) give the correct anomaly coefficients in the two cases.  This way of  
deriving the trace anomaly shows clearly that it is a scalar analog of the 
pseudoscalar axial-vector anomaly, a fact that will figure in the  
discussion of anomalies in supersymmetric theories.  

My involvement with trace anomalies began roughly five years later, when 
{\it Physical Review} sent me for refereeing a paper by Iwasaki [\refcite{iwas}].  In   
this paper, which noted the relevance to trace anomalies, Iwasaki proved 
a kinematic theorem on the vacuum to two photon matrix element of the  
trace of the energy-momentum tensor, 
that is an analog of the Sutherland-Veltman theorem for the vacuum to two  
photon matrix element of the divergence of the axial-vector current.  
Just as the latter has a kinematic zero at $q^2=0$, Iwasaki showed that 
the kinematic structure of the vacuum to two photon matrix element of the 
energy-momentum tensor implies, when one takes the trace, that there is 
also a kinematic zero at $q^2=0$, irrespective of the presence of   
anomalies (just as the Sutherland-Veltman result holds in the presence of 
anomalies).   Reading this article suggested the idea that just as the 
Sutherland-Veltman theorem can be used as part of an argument to prove 
nonrenormalization of the axial-vector anomaly, Iwasaki's theorem could be 
used to analogously calculate the trace anomaly to all orders.\footnote  
{In addition to writing a favorable report on Iwasaki's paper, I invited him to 
spend a year at the Institute for Advanced Study, which he did during the 
1977-78 academic year.}  During the spring of 1976    
I wrote an initial preprint attempting an all 
orders calculation of the trace anomaly in quantum electrodynamics,  
but this had an error pointed out to me by Baqi B\'eg.  
Over the summer of 1976 
I then collaborated with two local postdocs, John Collins (at Princeton) 
and Anthony Duncan (at the Institute), to 
work out a corrected version (Adler, Collins, and Duncan [\refcite{acd77}]). Collins 
and Duncan simultaneously teamed up with another Institute postdoc,  
Satish Joglekar, to apply similar ideas to quantum chromodynamics, published 
as Collins, Duncan, and Joglekar [\refcite{cdj77}], and simultaneously the same result 
for QCD was obtained by N. K. Nielsen [\refcite{nielsen}].  

In the simpler case of QED, a sketch of the argument based on Iwasaki's 
theorem goes as follows.                                                 
Let us write the trace of the energy-momentum tensor as 
\begin{equation}
\theta_{\mu}^{\mu}=K_1 m_0 \overline \psi \psi 
+ K_2 N[F_{\lambda \sigma}F^{\lambda \sigma}]+...,
\label{vox}
\end{equation} 
with ... denoting terms that vanish by the equations of motion, and with 
$ N[F_{\lambda \sigma}F^{\lambda \sigma}]$  specified by the conditions 
that its zero momentum transfer matrix elements between two electron 
states, and between the vacuum and a two photon state, are given by the 
corresponding tree approximation matrix elements of the 
operator $Z_3^{-1}F_{\lambda \sigma}F^{\lambda \sigma}$.   Taking 
matrix elements of $\theta_{\mu}^{\mu}$ between electron states at 
zero momentum transfer then gives the condition 
\begin{equation}
K_1\langle e(p)|m_0 \overline \psi \psi |e(p) \rangle =m~~~.
\label{wax}
\end{equation} 
However, an analysis using the Callan-Symanzik equation for the electron 
propagator shows that 
\begin{equation}
\langle e(p)|m_0 \overline \psi \psi |e(p) \rangle
={m \over 1+\delta(\alpha)}~~~,
\label{xyz}
\end{equation} 
with $1+\delta(\alpha)=(m/m_0)\partial m_0/\partial m$, so that $K_1$ is 
given by $K_1=1+\delta(\alpha)$.  Now taking the vacuum to two photon 
matrix element of $\theta_{\mu}^{\mu}$ at zero momentum transfer, and 
using Iwasaki's theorem and the defining condition for 
$ N[F_{\lambda \sigma}F^{\lambda \sigma}]$,  
we get a second condition 
\begin{eqnarray}
\label{yam}
0&=&[1+\delta(\alpha)]
\langle 0| m_0 \overline \psi \psi |
\gamma(p) \gamma(-p) \rangle \nonumber \\ 
&+&K_2 \langle 0 |Z_3^{-1}F_{\lambda \sigma}F^{\lambda \sigma} |
\gamma(p) \gamma(-p) \rangle_{\rm tree} ~~~.
\end{eqnarray}
However, analysis of the Callan-Symanzik equation for the photon 
propagator shows that 
\begin{eqnarray}
\label{zip}
\lefteqn{
\langle 0| m_0 \overline \psi \psi |
\gamma(p) \gamma(-p) \rangle  = -{1\over 4} {\beta(\alpha)\over 
1+ \delta(\alpha)} }\nonumber \\
&&\times   \langle 0 |Z_3^{-1}F_{\lambda \sigma}F^{\lambda \sigma} |
\gamma(p) \gamma(-p) \rangle_{\rm tree} ~~~,
\end{eqnarray}
with $\beta(\alpha)$ defined by $\beta(\alpha)= (m/\alpha) \partial \alpha/
\partial m$.  Hence $K_2$ is given by $K_2={1\over 4} \beta(\alpha)$, and  
so the final result for the trace equation is 
\begin{equation}
\theta_{\mu}^{\mu} = [1+\delta(\alpha)] m_0 \overline \psi \psi 
+ {1\over 4} \beta(\alpha)  N[F_{\lambda \sigma}F^{\lambda \sigma}]
+... ~~~.
\label{adj}
\end{equation}
The first two terms in the power series expansion of the coefficient 
of the $F_{\lambda \sigma}F^{\lambda \sigma}$ 
term in the trace agree with the fourth-order calculation of 
Chanowitz and Ellis.  The trace equation in QCD has a similar structure, 
again with the $\beta$ function appearing as the anomaly coefficient.

\section{Epilogue: Chiral and Trace Anomalies in Supersymmetric 
Yang-Mills Theories \label{epilogue}}

In Section~\ref{NCA} we saw that the chiral anomaly is given exactly by 
the one loop calculation, while in Section~\ref{trace} we saw that the trace anomaly  
can be calculated exactly in terms of the renormalization group 
$\beta$-function.  These two results have figured in a large literature 
dealing with anomalies in supersymmetric theories, and in particular  
in supersymmetric Yang-Mills theory, which is the case 
we shall focus on here (a similar analysis can be given when supersymmetric 
matter fields are included).    
We shall not attempt to cite all of the relevant papers, but note 
that reviews as of the mid-1980s were given 
in the comprehensive paper of Shifman and Vainshtein [\refcite{shifvain86}], and also 
in the introductory section of Ensign and Mahanthappa [\refcite{ensign87}].   Important 
recent discussions include Shifman's Sakurai Prize lecture [\refcite{shif99}] and 
the analysis of Arkani-Hamed and Murayama [\refcite{ahm}].  Although historically 
the topic of supersymmetric anomalies initially  generated much confusion, 
as I understand it the current status of the subject 
can be summarized as follows.  
\begin{enumerate}
\item In supersymmetric Yang-Mills theory at the {\it classical} level, 
there is a supercurrent containing among its components both the axial-vector  
current and the energy momentum tensor (Ferrara and Zumino [\refcite{fezu}]; 
Piguet and Sibold [\refcite{ps82a}]).  Thus, at the {\it quantum} level, where 
anomalies appear, one might expect there to be a relation between the 
axial-vector anomaly and the trace anomaly 
(Grisaru [\refcite{gris79}]; Piguet and Sibold [\refcite{ps82b}]). In fact, since we have seen that 
the scalar trace anomaly is an analog of the pseudoscalar chiral anomaly, and 
since chiral supermultiplets in supersymmetry contain scalar and pseudoscalar 
fields on an equal footing, we should expect there to be an anomaly chiral 
supermultiplet.\footnote{All components of the supercurrent have anomalies.  For example, the one-loop anomaly in the spinor supersymmetry current was obtained by Curtright [\refcite{curt}], and the one-loop supercurrent anomalies for the Wess-Zumino model were obtained by Clark, Piguet, and Sibold [\refcite{cps1}]. For a summary of the one-loop supercurrent anomalies in component field form, and further references, see Appendices B and C of Adler [\refcite{asq}] and Chapter 20 of West [\refcite{we}].} 
However, even in the supersymmetric case there is an axial-vector current 
that obeys the Adler-Bardeen theorem (Clark, Piguet, and Sibold [\refcite{cps2}]; Jones and Leveille [\refcite{jlev82}]; 
Piguet and Sibold [\refcite{ps86}]; Ensign and 
Mahanthappa [\refcite{ensign87}]), whereas a number of calculations have shown that there 
are higher loop contributions to the trace anomaly (Jones [\refcite{jdrt83}];  
Novikov, Shifman, Vainshtein, and Zakharov [\refcite{nsvz83}]; Grisaru and West [\refcite{gwest}]).  
These authors gave differing arguments (with differing 
degrees of validity --  
see Novikov, Shifman, Vainshtein, and Zakharov  [\refcite{nsvz85}]) 
for what is now called the NSVZ $\beta$ function, given by Eq.~(\ref{ban}) below. 
Thus, there seems to be a potential conflict 
between supersymmetry and the all orders calculations of anomalies.   

\item The resolution to the potential conflict consists in noting 
that there is more than one consistent choice of subtraction scheme for the
renormalized currents and their anomalies (Jones, Mezincescu, and West [\refcite{jmw85}];  
Grisaru, Milewski, and Zanon [\refcite{gmz85,gmz86}]).  Thus, there are two manifestly 
supersymmetric supercurrents, which coincide at the classical level, 
one of which satisfies the Adler-Bardeen 
theorem, with a one-loop anomaly, and the other which has an anomaly given 
by the multiloop $\beta$-function.  The paper of Jones, Mezincescu, and West 
shows that the relation between the two supercurrents involves two functions 
of coupling which they call $a(g)$ and $b(g)$; when one assumes that these 
two functions are equal one can derive a formula for the $\beta$ function 
to all orders, 
\begin{eqnarray}
\label{ban}
\beta(g)&=&{\beta^{(1)}(g) \over 1+ {2\over 3} \beta^{(1)}(g)/g }~~~,\nonumber \\
\beta^{(1)}(g)&=&-3C_2(G)g^3/(16 \pi^2)~~~,
\end{eqnarray}
with $\beta^{(1)}$ the one-loop $\beta$ function.  The result of Eq.~(\ref{ban})  
is what was found in the instanton calculation of Novikov, Shifman, 
Vainshtein, and Zakharov  [\refcite{nsvz83}] and also in an earlier calculation of 
Jones  [\refcite{jdrt83}]; however, this earlier Jones result cannot be regarded as 
an alternative derivation of the NSVZ $\beta$ function, since it has 
implicitly assumed $a(g)=b(g)$.   

\item  We note that if it were not for the 
denominator  $1+ {2\over 3} \beta^{(1)}(g)/g$, Eq.~(\ref{ban}) would be saturated  
by the one-loop result.  Thus Eq.~(\ref{ban}) can be regarded as a  
renormalization 
of a one-loop result, and as such, it is simpler in structure than the  
$\beta$ functions found in the non-supersymmetric case, where the chiral   
and trace anomalies are unrelated (witness the factor of $1/3$ in the trace 
anomaly coefficient) even in leading one-loop order.  This is the 
basis for the modern interpretation of the supersymmetric ``anomaly puzzle''.  
According to this interpretation, first proposed by 
Shifman and Vainshtein  [\refcite{shifvain86}], the supercurrent generalizing the 
Adler-Bardeen   
theorem has an anomaly given by the one loop term in Eq.~(\ref{ban}), whereas the 
supercurrent generalizing the trace anomaly discussed in Section~\ref{trace} has 
the $\beta$ function of Eq.~(\ref{ban}), including the renormalizing denominator.  
These two differing supercurrents in fact correspond to two 
different choices of coupling constant appropriate to two different 
calculational schemes, with the renormalizing denominator arising from the    
transformation between the two calculational schemes.  
If one uses a Wilsonian effective action, 
in which only virtual momenta greater than some (large) minimum $\mu$ are 
kept, then the supercurrent anomaly is exhausted at one loop order.  On 
the other hand, if one uses a one particle irreducible effective action 
with a canonically normalized coupling constant, the associated coupling 
renormalization leads to a supercurrent anomaly given by the NSVZ $\beta$ 
function of Eq.~(\ref{ban}).  

\item Arkani-Hamed and Murayama  [\refcite{ahm}] have given a very interesting 
interpretation of the two possible supercurrents and anomalies in terms 
of how one introduces the bare coupling in the Lagrangian, working entirely   
within a Wilsonian action framework, without reference to the one 
particle irreducible effective action invoked by Shifman and Vainshtein.  
\big(For a related   
commentary, see Shifman  [\refcite{shif99}].\big) If one uses 
a manifestly holomorphic coupling definition 
\begin{equation}
{\cal L}={1\over 16} \int d^2\theta {1\over g_h^2} [W^a(V)]^2  + ~
{\rm adjoint}~~~,
\label{cup}
\end{equation} 
with $1/g_h^2=1/g^2 + i \theta/(8\pi^2)$, and $V=V^{\dagger}$ the vector 
superfield, holomorphicity arguments of the 
type used to prove supersymmetry nonrenormalization theorems show that 
the $\beta$ function is exhausted at one-loop order, in agreement with   
the Shifman-Vainshtein construction based on the Wilsonian effective action. 
This corresponds to 
the definition of the supercurrent multiplet that contains an axial-vector 
current obeying the Adler-Bardeen theorem.  On the other hand,  one can 
instead use  
a ``canonical'' coupling, 
\begin{equation}
{\cal L}={1\over 16} \int d^2 \theta\left( {1\over g_c^2} +i 
{\theta \over 8 \pi^2} \right) [W^a(g_c V)]^2~~~,
\label{dax}
\end{equation}  
defined as one that gives 
the kinetic energy a coupling-independent coefficient.  This is  
not holomorphic in $g_c$ because of the self-adjoint restriction on the 
superfield argument $g_cV$ of the field-strength superfield $W^a$.  
Since transforming from the holomorphic coupling to the canonical coupling 
involves a field rescaling, there is a Jacobian associated with this 
transformation, that can be computed in analogy with the non-supersymmetric 
Fujikawa  [\refcite{fujik80b}] trace anomaly calculation 
and the later supersymmetric Konishi-Shizuya  [\refcite{kksk85}] anomaly calculation.  
Arkani-Hamed and Murayama show that this Jacobian precisely accounts for 
the difference between the one loop $\beta$ function and the NSVZ $\beta$ 
function of Eq.~(\ref{ban}), corresponding to the transformation given by   
Shifman and Vainshtein,
\begin{equation}
{1\over g_c^2}={\rm Re}\left({1\over g_h^2}\right) -{2C_2(G) \over 8\pi^2} 
\log g_c~~~,
\label{esp}
\end{equation} 
from the Wilsonian to the one particle irreducible 
effective action.  A standard result of the renormalization group states   
that under analytic transformations of the coupling constant, the one- and 
two-loop $\beta$ functions are invariant, with changes appearing only at 
three- and higher loop order.  However, because 
the transformation of Eq.~(\ref{esp}) has a non-analytic, logarithmic dependence on 
$g_c$, it can transform the one-loop beta function $\beta^{(1)}$ 
into the NSVZ $\beta$ function, which has a nonvanishing two-loop term: letting $\Lambda$ be the cutoff, differentiating Eq.~(\ref{esp}) implies that
\begin{equation}
{\beta(g_{c})\over \beta(g_{h})} = {\Lambda dg_{c}/d \Lambda \over \Lambda dg_{h}/d \Lambda} = 
{dg_{c} \over dg_{h}} = {g_c^3 \over g_h^3 } \left( 1 - {C_2(G) \over 8 \pi^2} g_c^2 \right)^{-1}~~~,
\label{flag}
\end{equation}
which with $\beta(g_{h})= -3C_2(G)g_h^3/(16 \pi^2)$ implies Eq.~(\ref{ban}) for $\beta(g_{c})$.  Although we have seen that without further input, functional integral 
Jacobian calculations may give only leading order results, 
Arkani-Hamed and Murayama give a second method of indirectly 
determining the Jacobian by using the finiteness of $N=2$ supersymmetric 
theories, which shows that the one-loop answer for the 
Jacobian is exact, a conclusion also reached in the earlier 
paper of  Shifman and Vainshtein.   
This thus closes the gap of showing that $a(g)=b(g)$ left 
unresolved in the work of Jones, Mezincescu, and West, and gives    
an alternative all orders derivation of the NSVZ result, and explains in 
a simple way why   
the NSVZ formula has the structure of a one-loop $\beta$ function 
up to a renormalizing factor.   
\end{enumerate}

To sum up, there now appears to be a detailed understanding of the 
role played by all orders calculations of chiral and trace 
anomalies in supersymmetric gauge theories.  In retrospect, the 
``supersymmetric anomaly puzzle'' stemmed from the initial, preliminary  
generalization from 
known non-supersymmetric results to the supersymmetric case,  
which suggested that one should find a unique supercurrent 
with a one-loop anomaly structure.  
What one finds, instead, is an equivalence class of supercurrents, differing   
by the coupling constant definition, with a one-loop anomaly structure 
{\it modulo renormalizations  arising from 
redefinitions of the coupling constant} -- different from first expectations,   
but not that far off.  This contrasts with the all orders  
anomalies found in QED and QCD, where the trace anomaly is given by a 
$\beta$ function that is unrelated to the chiral anomaly even at 
one-loop order, reflecting the 
fact that these theories are not supersymmetric.

To conclude, I note that I have only touched in this essay on one part 
of a large subject.  For further, detailed discussions of anomalies, 
see the books of van Nieuwenhuizen  [\refcite{vnp88}] and Bertlmann  [\refcite{bert}], as well 
as  chapters or sections of the books by Weinberg  [\refcite{wein96}], Makeenko [\refcite{makeen}],  
Volovik [\refcite{vol}], and Zee  [\refcite{zee03}], and of a forthcoming book 
by van Nieuwenhuizen.\footnote{P. van Nieuwenhuizen, {\it Advanced Quantum 
Field Theory} (in preparation).}

\section*{Acknowledgments}
This work was supported in part by the Department of Energy under
Grant \#DE--FG02--90ER40542.   I wish to thank Bill Bardeen, 
John Collins, Rod  
Crewther, and Tony Zee for helpful conversations and/or email correspondence  
on the sections relating to their work, Serge Rudaz for bringing a pertinent 
reference to my attention,  Hitoshi Murayama for a critical 
reading of Section~\ref{epilogue}, and Peter van Nieuwenhuizen for a critical reading 
of the entire manuscript and many useful suggestions.

\appendix
\section{Some Details of the Zumino 
and Schwinger Calculations}

I give here a few key formulas from the Zumino  [\refcite{zumino69}] point-splitting 
calculation, that allow one to understand what is being done in the   
Schwinger  [\refcite{sch51}] calculation.  I follow the notation used by both 
Zumino and Schwinger, and 
so in this Appendix $A_{\mu}$ denotes the electromagnetic vector potential, 
that was denoted by $B_{\mu}$ in the text, and the metric conventions also 
differ from those used in the text.  Zumino's treatment starts from 
the formula
\begin{equation} 
[\partial_{\mu}-ieA_{\mu}(x)
+\partial_{\mu}^{\prime} + ieA_{\mu}(x^{\prime})] {\rm tr}\gamma_5 
\gamma^{\mu} G(x,x^{\prime})=-2m{\rm tr} \gamma_5 G(x,x^{\prime})~~~,
\label{fall}
\end{equation}
which is immediately obtained 
from the electron Dirac equation in the presence 
of an electromagnetic field,   
and the definition of the electron Green's function $G$.  
Zumino defines a gauge invariant axial-vector current 
by the $x \to x^{\prime}$ limit of the point-split expression   
\begin{equation}
{\rm tr}\gamma_5 \gamma_{\mu} G(x,x^{\prime}) 
\exp[-ie\int_{x^{\prime}}^x A \cdot dx]~~~. 
\label{gnome}
\end{equation} 
(Note that this is opposite in sign to the axial-vector current  
used above in the text, because both Zumino and Schwinger define the  
axial-vector current with the ordering $\gamma_5 \gamma_{\mu}$, whereas 
in the text we use the ordering $\gamma_{\mu} \gamma_5$.) 
To calculate the divergence of this formula, he uses the easily derived 
identity 
\begin{eqnarray}
\label{hat}
&[\partial_{\mu}-ieA_{\mu}(x)
+\partial_{\mu}^{\prime} + ieA_{\mu}(x^{\prime})] 
\exp[ie\int_{x^{\prime}}^x A \cdot dx] \nonumber \\
=&\exp[ie\int_{x^{\prime}}^x A \cdot dx]
[\partial_{\mu}+\partial_{\mu}^{\prime}+ie(x-x^{\prime})^{\nu} F_{\mu\nu}]~~~,
\end{eqnarray}
with $F_{\mu\nu}=\partial_{\mu}A_{\nu}-\partial_{\nu}A_{\mu}$ the 
field strength evaluated at the midpoint of the 
Wilson line.  Rewriting this formula as 
\begin{eqnarray}
\label{ing}
&\exp[-ie\int_{x^{\prime}}^x A \cdot dx]  [\partial_{\mu}-ieA_{\mu}(x)
+\partial_{\mu}^{\prime} + ieA_{\mu}(x^{\prime})] 
 \nonumber \\
=&[\partial_{\mu}+\partial_{\mu}^{\prime}+ie(x-x^{\prime})^{\nu} F_{\mu\nu}]
\exp[-ie\int_{x^{\prime}}^x A \cdot dx] ~~~,
\end{eqnarray}
one finds by use of Eqs.~(\ref{fall}) and (\ref{gnome}) that  
\begin{eqnarray}
\label{jump}
&[\partial_{\mu}+\partial_{\mu}^{\prime}+ie(x-x^{\prime})^{\nu} F_{\mu\nu}]
{\rm tr}\gamma_5\gamma^{\mu}G(x,x^{\prime}) 
\exp[-ie\int_{x^{\prime}}^x A \cdot dx] \nonumber \\
&=-2m{\rm tr}\gamma_5G(x,x^{\prime}) \exp[-ie\int_{x^{\prime}}^x A \cdot dx] ~~~.
\end{eqnarray}
The right hand side of Eq.~(\ref{jump}) is the point-split definition of the 
naive divergence.  The left hand side is a sum of the point-split definition 
of the ordinary divergence of the gauge-invariant point-split axial-vector 
current, and an extra term that gives the anomaly.  The anomaly term is 
readily evaluated by using the formula given by Schwinger and  
Zumino for the small $x-x^{\prime}$ 
behavior of the Green's function, 
\begin{equation}
{\rm tr} \gamma_5\gamma_{\mu} G(x,x^{\prime}) 
   \sim {ie \over 2\pi^2} \hat F_{\mu\lambda} {(x-x^{\prime})^{\lambda} 
\over (x-x^{\prime})^2}
\exp[ie\int_{x^{\prime}}^x A \cdot dx]~~~, 
\label{kab}
\end{equation}
with $\hat F_{\mu\lambda}={1\over 2}\epsilon_{\mu\lambda\rho\sigma}F^{\rho
\sigma}$ the dual electromagnetic field strength.   

Schwinger's calculation consists of an evaluation of the left hand side 
of Eq.~(\ref{jump}) by first using Eq.~(\ref{kab}) \big(which is derived in Schwinger's 
Eqs.~(5.17)-(5.21)\big), and then using Eq.~(\ref{hat}) \big(which is applied in 
Schwinger's Eq.~(5.22)\big).  As noted in the text, 
Schwinger's identification of a gauge-covariant axial-vector divergence 
is incorrect -- apart from the overall minus sign noted above, 
the object that he so labels is actually (up to a Wilson  
line factor) the gauge-invariant axial-vector divergence, minus 
the axial-vector anomaly term.  
There is no gauge-invariant 
axial-vector current for which this combination is the divergence, but 
as shown in Eqs.~(58) and (59) of 
Adler  [\refcite{adler69a}], 
one can readily write down a gauge-non-invariant axial-vector current 
which has this divergence.


\end{document}